\RequirePackage{fix-cm}
\documentclass[smallextended]{svjour3}       % onecolumn (second format)
\smartqed  % flush right qed marks, e.g. at end of proof
\usepackage{graphicx}

\begin{document}

\title{Analysis of temporal properties of wind extremes}


\author{Luciano Telesca       \and
        Fabian Guignard \and
        Mohamed Laib  \and
        Mikhail Kanevski
}


\institute{L. Telesca \at
              Istituto di Metodologie per l'Analisi Ambientale, CNR \\
              C.da S.Loja, 85050 Tito (PZ), Italy.
 \\
              \email{luciano.telesca@imaa.cnr.it}           %  \\
           \and
           F. Guignard, M. Laib, and M. Kanevski \at
              IDYST, Faculty of Geosciences and Environment \\
              University of Lausanne 1015, Switzerland. \\
              Corresponding author: (M. Laib) \\
              \email{Mohamed.Laib@unil.ch}  
}

\date{Received: date / Accepted: date}

\maketitle

\begin{abstract}
The 10-minute average wind speed series recorded at 132 stations distributed rather homogeneously in the territory of Switzerland are investigated. Wind extremes are defined on the base of run theory: fixing a percentile-based threshold of the wind speed distribution, a wind extreme is defined as a sequence of consecutive wind values (or duration of the extreme) above the threshold. This definition allows to analyse the sequence of extremes as a temporal point process marked by the duration of the extremes. The average probability density function of the duration of the extremes of the wind speed measured in Switzerland does not depend on the percentile-based threshold and decrease with the increase of the extreme duration. The time-clustering behaviour of the sequences of the wind extremes was analysed by using the global and local coefficient of variation and the Allan Factor. The wind extremes are globally time-clustered, although they tend to behave as a Poisson process with the increase of the minimum extreme duration. Locally, the wind extremes tend to be clustered for any percentile-based threshold for stations located above about 2,000 m a.s.l. By using the Allan Factor, it was revealed that wind extremes tend to be clustered even at lower timescales especially for the higher stations.
\end{abstract}

\section{Introduction}
\label{intro}
Among the different factors to consider in climate dynamics, wind is surely an important one. Wind speed research has impacted on several fields, like energy generation \cite{ref1}, air pollution control \cite{ref2}, civil engineering \cite{ref3}, aeolian sediment transport \cite{ref4}, to mention few of them. Near-surface wind fluctuations show very irregular and complex behaviour, due to several interacting factors, such as pressure gradient, turbulence phenomena, temperature, morpho-topographic conditions \cite{ref5}. The traditional simulation models, like wind tunnel simulations \cite{ref6} or computational ﬂuid dynamics methods \cite{ref7} were quite limited in revealing the dynamical complexity of wind field. Therefore, the dynamical characterization of wind speed series by using robust methods has been focus of several researches, from the distributional analysis \cite{ref8} 
to chaotic time series analysis\cite{ref9}, 
%from wavelet [10] and spectral analysis\cite{ref11} 
and fractal/multifractal analysis \cite{ref14} \cite{ref15} \cite{ref16} \cite{ref17} \cite{ref18} \cite{ref19} \cite{ref20}, from multiscale entropy analysis \cite{ref13} to multiscale multifractal analysis \cite{zeng2016}. The most of the wind studies dealt with hourly or daily means of wind speed. High-frequency wind records would be needed to capture inner features of its dynamics, like turbulence phenomena \cite{ref23} and to disclose complex dynamical patterns at timescales lower than those that are being generally investigated. 

A special focus of climate studies are the extremes, which are values that depend mainly on the definition of the distribution and on how far into the tail of the distribution the threshold value is located. Generally, values located in the far tail of the distribution tend to be more crucial to societal and natural systems than values occurring more frequently. In fact the more extreme an event is, the more probably larger the damage is, that can produce to the society and environment. 
As it was observed by Zhang et al. \cite{Zhang2011}, analyzing changes of the frequency or intensity of extremes, which are further out in the tail of the distribution leads to the results that are inherently more uncertain because of their smaller occurrence frequency. For instance, using only 50 years of daily data, it is a hard task to  estimate the frequency changes of events with a 50-year return period. Percentile-based extremes are frequently used in the analysis of climate extremes because they provide sufﬁciently large sample sizes for robust statistical assessments \cite{Martius2016}. In hydrology, for instance, Froidevaux et al. \cite{Froidevaux2015} showed for Swiss catchments that more than $20\%$ of all ﬂood events with a 5 year return period are preceded by 2 day precipitation sums between the 95th  and the 99th  percentile. Klawa and Ulbrich \cite{Klawa2003} analyzed large-scale windstorms affecting large areas and producing large loss. Since  each station is characterized by its particular wind climate (high wind speeds frequently occur on an exposed areas without causing any losses, while same wind speeds can produce large loss in other areas, with different topographic conditions). They proposed to normalize wind speed with a local climatological extreme wind speed to remove the effect of differences in wind climate and allow to spatially interpolate normalized station wind speeds. In their analysis, it was  found that a wind threshold of the 98th percentile of the wind speed distribution could be sufficiently relevant for taking into account of possible  damages produced by windstorms. Martius et al. \cite{Martius2016} analysed the co-occurrence of wind and precipitation extremes, both defined as the values above the 98th percentile. 

In this work, we investigate the high-frequency wind speed data measured by the MeteoSwiss weather network at 132 stations widespread all over the territory of Switzerland from 2008 to 2017 (Fig. 1 shows a map of Switzerland with the  stations). The data were sampled at 10 min and a deep analysis of the temporal properties of wind extremes will be performed. In particular the run theory will be used to define the extremes and a time-clustering analysis, by using several methods, will be carried out to evidence correlation structures in the sequence of wind extremes. These approaches have never been applied to wind, up to our knowledge.  
\begin{figure}
%%\rule{1cm}{1cm}width=\linewidth
\includegraphics[width=\linewidth]{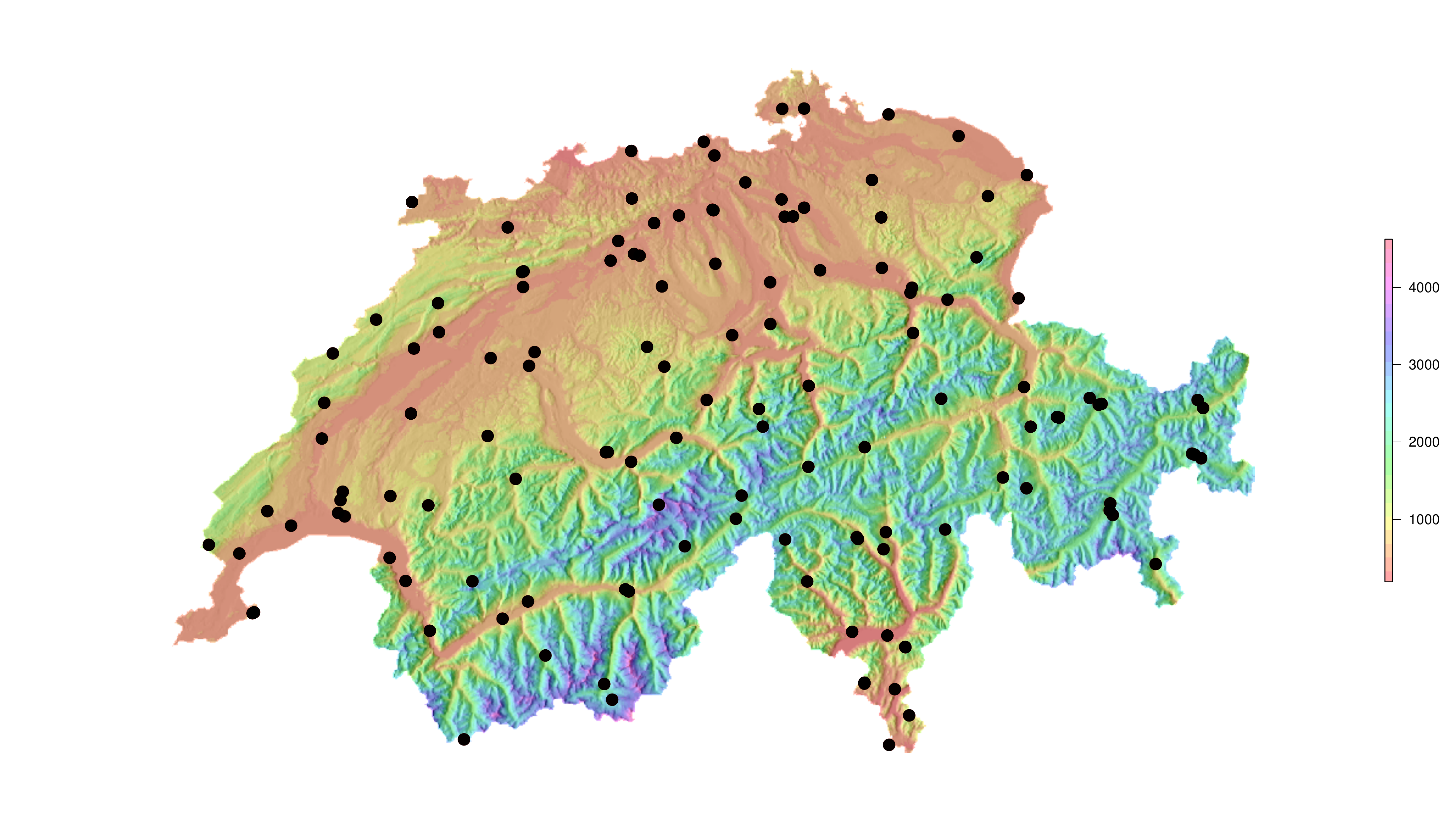}
\caption{Study area and location of the measurement stations.}
\label{fig1}  
\end{figure}

\section{Definition of extremes in wind and their statistical features}
Using the percentile-based definition of wind extreme, in our study three thresholds were investigated: 95th, 97.5th and 99th percentile of the wind speed distribution of each station.  Fig. 2 shows the relationship between the three wind values corresponding to the percentile thresholds of $95\%$, $97.5\%$ and $99\%$ thresholds and the height of the station.  Although a slight increase of the threshold with the height is apparent, it seems also evident a height crossover at about $2000$ m a.s.l. separating the wind thresholds.

\begin{figure}
%%\rule{1cm}{1cm}width=\linewidth
%\centering
\includegraphics[width=\linewidth]{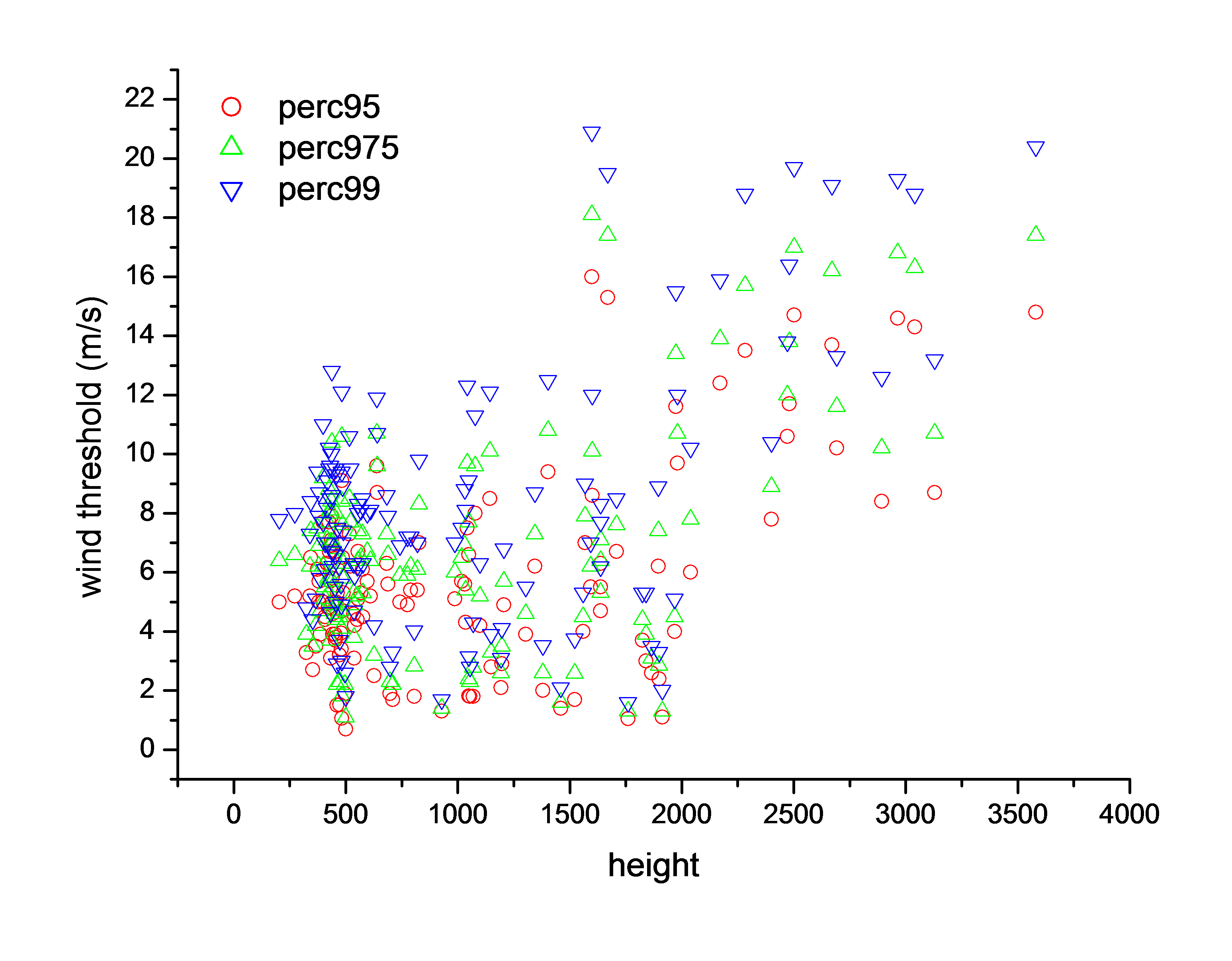}
\caption{Wind value versus height for different percentile-based thresholds.}
\label{fig2}  
\end{figure}

On the basis of the crossing theory or run theory \cite{Harald1967}, we define a "run" as a sequence of contiguous values above a given percentile-based threshold. A characteristic of a run is its length that represents the period m in which the variable under study is above the selected threshold. Hereafter, we intend as wind extreme a run with a specific length m. 

A sequence of runs or extremes can be viewed as a point process in time. Indicating with $t_i$ the time at which the run starts or the extreme event occurs  (hereafter, time of the run) and with $L_i$ its length or duration of the extreme, the sequence of extremes can be described as a finite sum of Dirac’s delta functions centred on the time $t_i$ and amplitude proportional to $L_i$:

\begin{equation}
y(t)=\sum_{i=1}^{N}L_i \delta(t-t_i)
\end{equation}
where $N$ is the length of the sequence of extremes. Fig. 3 shows, as an example, the sequence of wind extremes (for the threshold of 95th percentile of the distribution of the wind speed) at the station WSLVSF at $640$ m a.s.l.

\begin{figure}
%%\rule{1cm}{1cm}width=\linewidth
%\centering
\includegraphics[width=\linewidth]{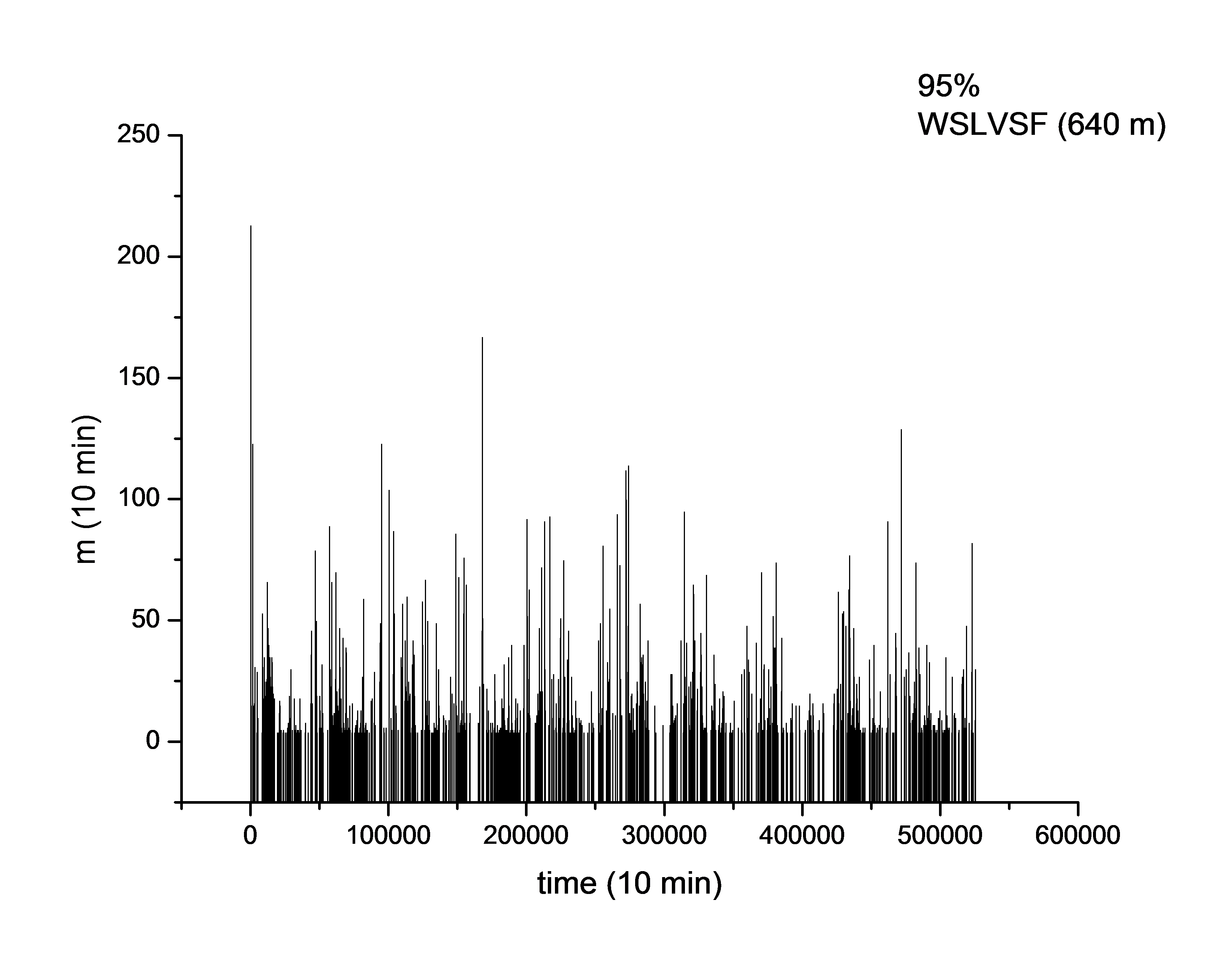}
\caption{Sequence of extremes for the station WSLVSF with the percentile-based threshold of $95\%$ .}
\label{fig3}  
\end{figure}

A statistically relevant quantity is the  probability density function $P(m)$ of the run length or extreme duration m; we calculated $P(m)$ for each of the three thresholds, along with their average $<P(m)>$ (Fig. 4).

\begin{figure}
\begin{tabular}{ll}

\includegraphics[scale=0.1]{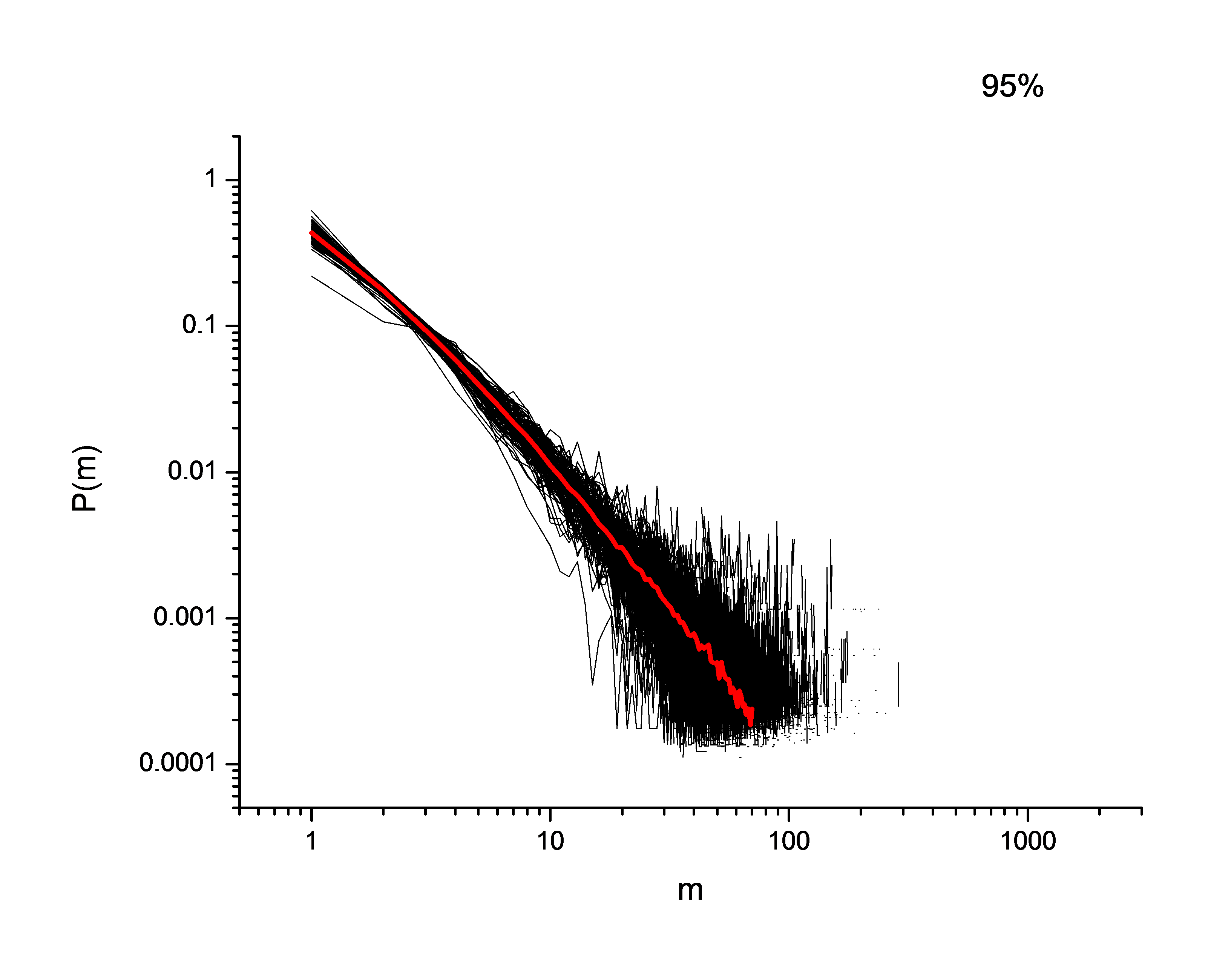} &

\includegraphics[scale=0.1]{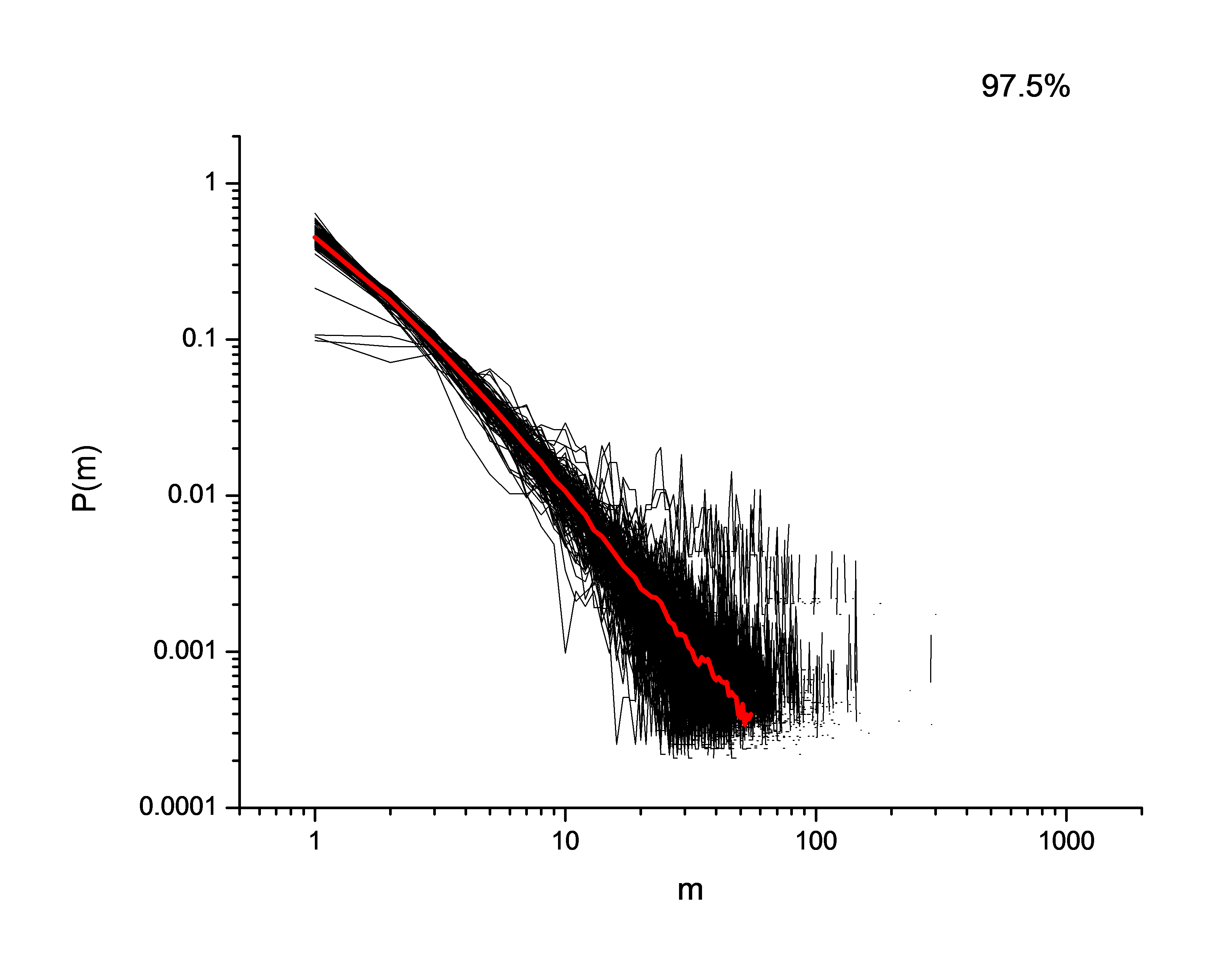}   \\

\includegraphics[scale=0.1]{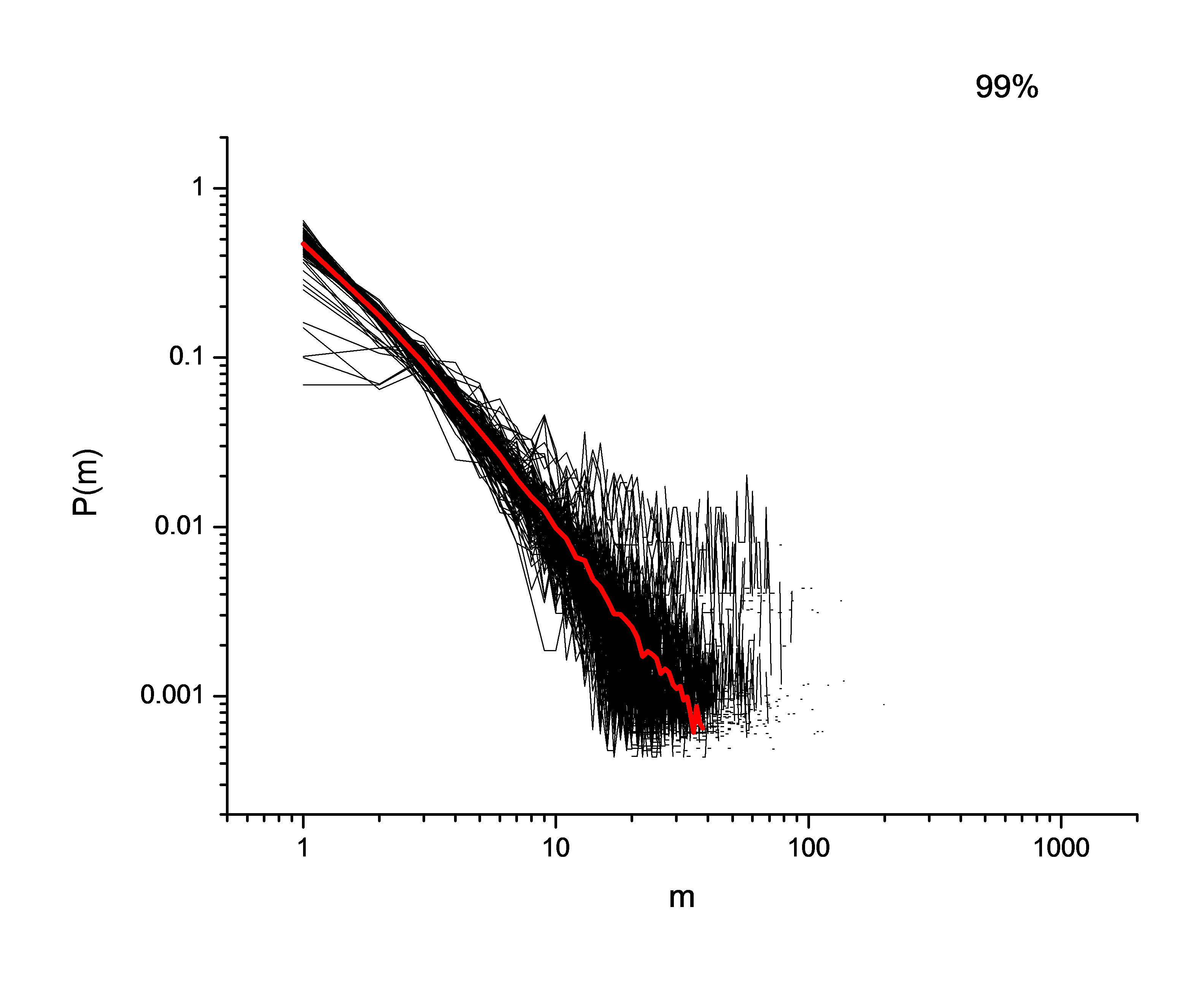}  &

\includegraphics[scale=0.1]{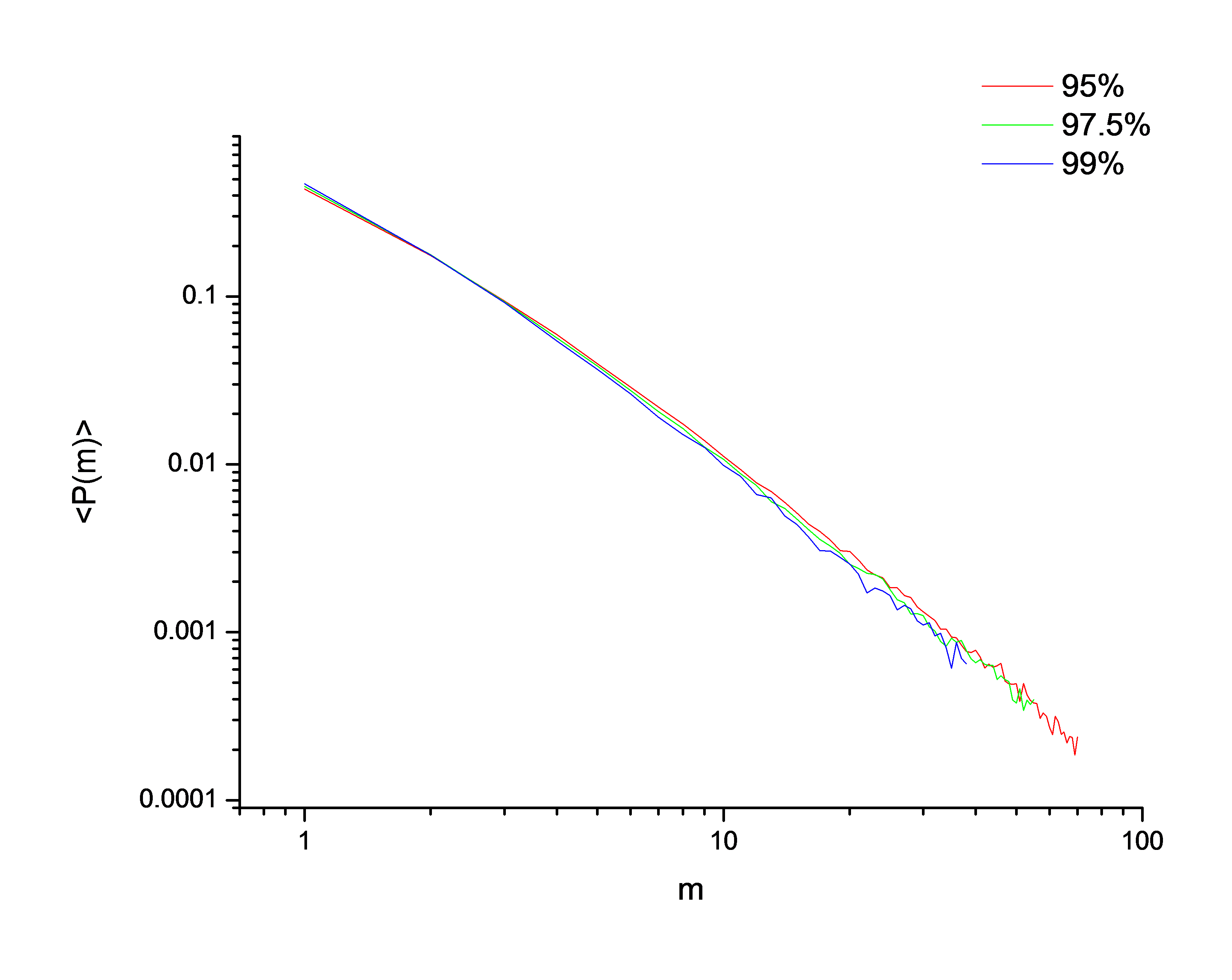}   \\

\end{tabular}
\caption{$P(m)$ for percentile-based thresholds of $95\%$ (top-left panel), $97.5\%$ (top-right panel) and $99\%$ (bottom-left panel). (bottom-right panel) Comparison among the $<P(m)>$ for each threshold.}
\label{fig4}  
\end{figure}

It is visible that $P(m)$ roughly decreases with $m$ for any station; a specific relationship with the height cannot be observed.  In fact, for instance, $P(m)$ of the stations AIG at $381$ m a.s.l. and ARD at $1840$ m a.s.l. do not show striking difference (Fig. 5a), while $P(m)$ of station WSLVOB at $482$ m a.s.l. is very different from that of station WSLVOF at $480$ a.s.l. (Fig. 5b).

\begin{figure}
\begin{tabular}{l}

(a)

\includegraphics[scale=0.1]{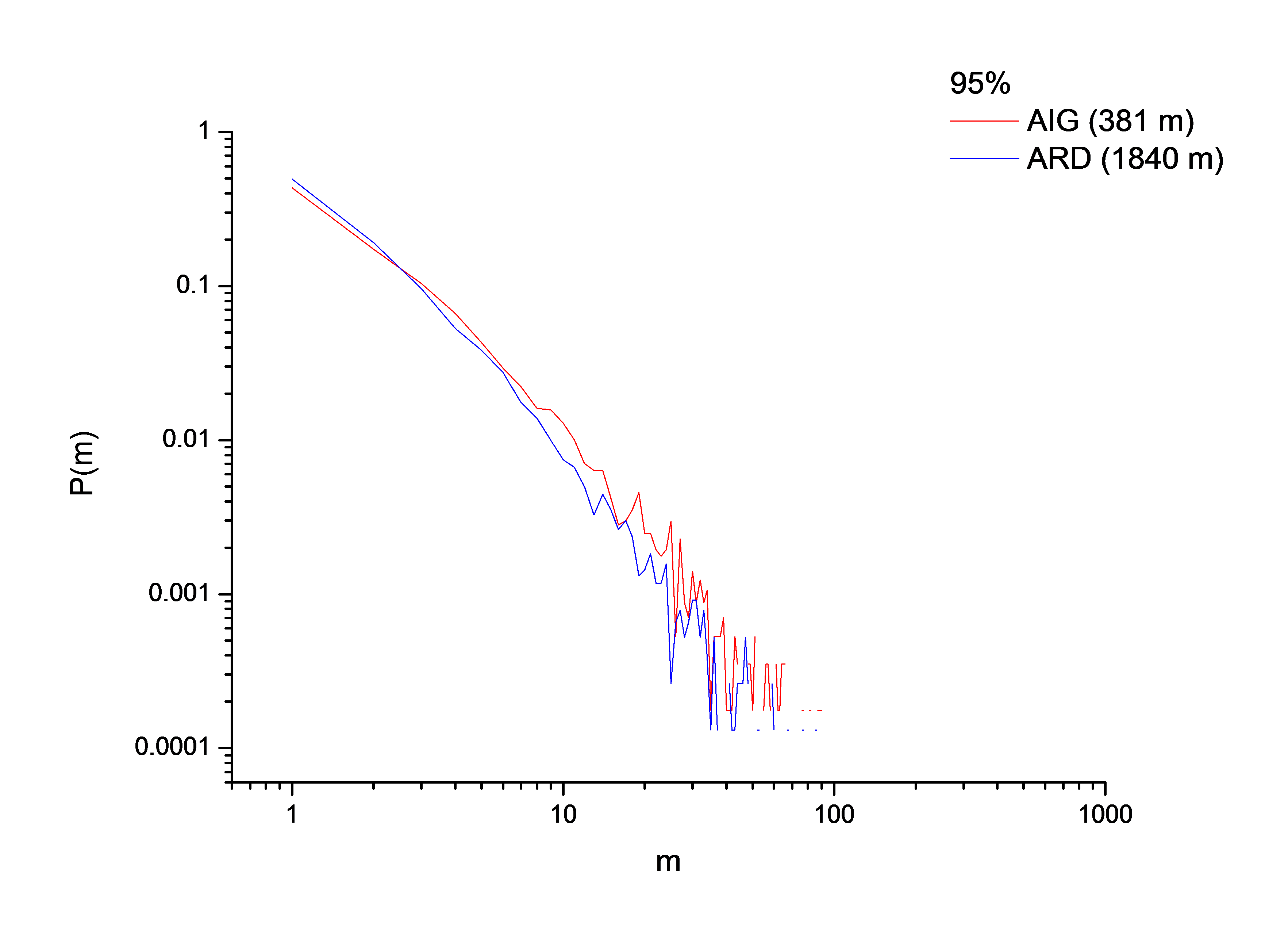}   \\

(b)

\includegraphics[scale=0.1]{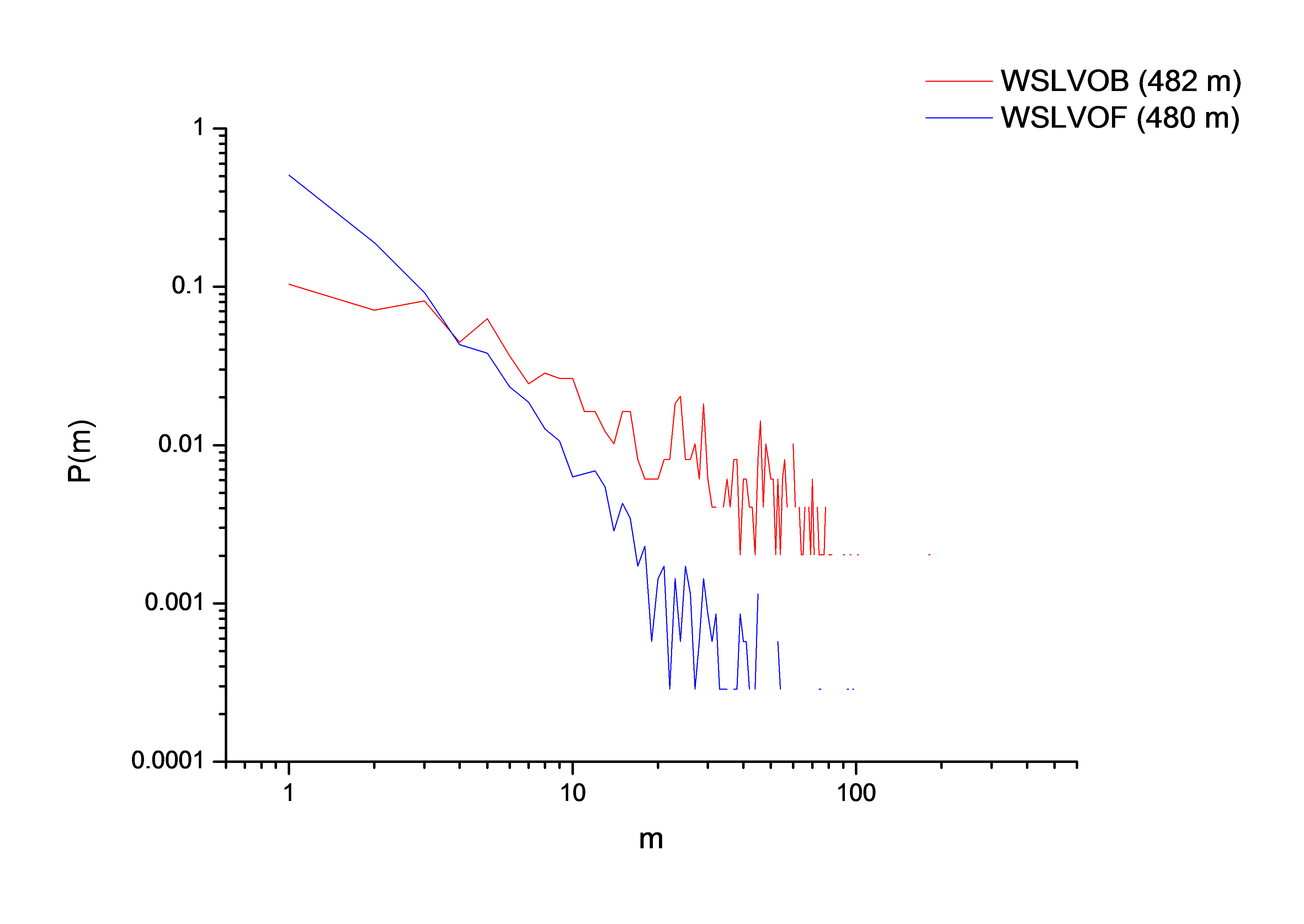}   \\

\end{tabular}
\caption{Comparison between $P(m)$ of (a) AIG and ARD at different heights of $381$ m a.s.l. and $1840$ m a.s.l. respectively, and (b) between WSLVOB and WSLVOF at about the same altitude.}
\label{fig5}  
\end{figure}

Averaging $P(m)$ for each threshold (red curve in Fig. 4a,b,c), and comparing them (Fig. 4d), we see a rather identical behaviour, indicating that $<P(m)>$ does not depend on the threshold used to define the extremes. 

\section{Time-clustering of wind extremes}

The sequence of wind extremes can be represented by interevent times or by  counting processes. In the first case, a discrete-time series is formed by the rule $T_i=t_{i+1} - t_i$ , where $t_i$ indicates the time of the extreme numbered by the index $i$. In the second case, the time axis is divided into equally spaced contiguous counting windows of duration $\tau$ to produce a counting process $\{Nk(\tau)\}$, where $Nk(\tau)$ represents the number of extremes falling into the $k-th$ window of duration $\tau$. The duration $\tau$ of the window is called timescale. The latter approach considers the extremes as the events of interest assuming the existence of an objective clock for the timing of the events. The first approach emphasizes the interevent intervals using the event number as an index of the time. Fig. 6 shows, as an example, the interevent time series of the station WYN at $422$ m a.s.l. for the threshold of $95\%$.

\begin{figure}
%%\rule{1cm}{1cm}width=\linewidth
%\centering
\includegraphics[width=\linewidth]{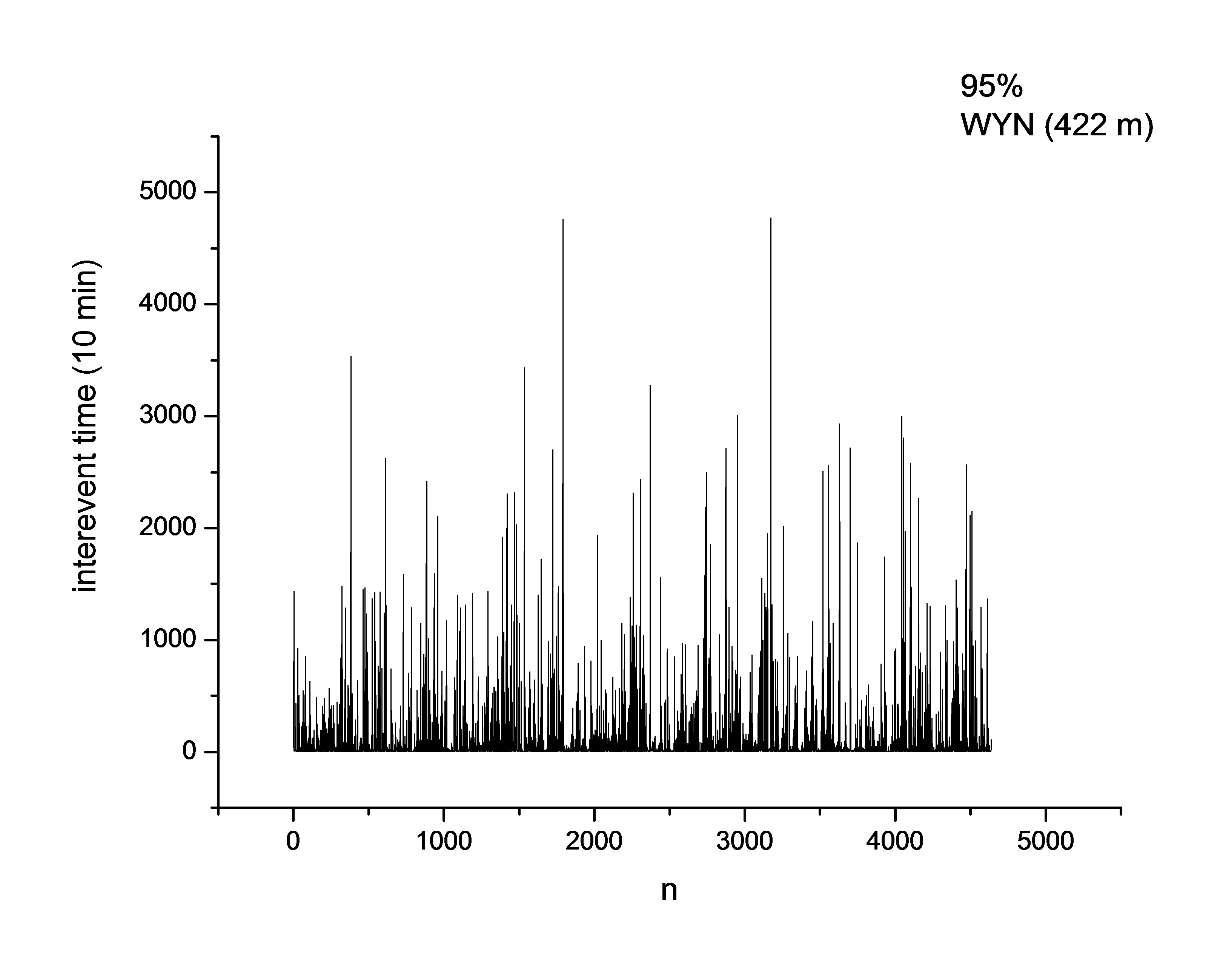}
\caption{Interevent time series of the wind extremes (percentile-based threshold of $95\%$) at station WYN, located at $422$ m a.s.l. }
\label{fig6}  
\end{figure}

Depending on the representation, various statistical quantities can be defined and used to quantify the time-clustering: the coefficient of variation, global ($C_v$) and local ($L_v$) for the interevent times and the Allan Factor (AF) for the counting processes.

\subsection{Coefficient of variation}
The global coefficient of variation ($C_v$)  \cite{Kagan1991} is defined as
\begin{equation}
C_v= \sigma_T /<T>
\end{equation} 
where $<T>$ and  $\sigma_T$ are the average and the standard deviation of the interevent times: a Poisson process (memoryless, completely random) is characterized by  $C_v=1$, a periodic or regular process by $C_v < 1$, but a clustered process by $C_v>1$. This coefficient allows to evaluate globally if a process is clustered or not, although it does not furnish information about the timescale ranges where clustering appears. This could be a limit of this measure, since the deep comprehension of a complex process, like wind extremes, could be possible if the different timescale ranges governing its dynamics are well detected. Recently, Telesca et al. \cite{Telesca2016} investigated the time-clustering properties of the point process of volcanic seismicity at El Hierro, Canary Islands (Spain) by using the local coefficient of variation ($L_v$), defined by Shinomoto et al. \cite{SHINOMOTO200567} as:
\begin{equation}
L_v=\frac{1}{N-1}\sum_{i=1}^{N-1}3 \frac{(T_i- T_{i+1})^2}{(T_i + T_{i+1})^2}
\end{equation}
where $N$ is the number of interevent times. Like the $C_v$, $L_v$ is larger, equal or smaller than $1$ if the process is clustered, Poissonian or regular. Contrarily to  $C_v$, $L_v$ describes locally the variability of the interevent times. For instance, if a point process is a combination of two periodic sequence, $C_v>>1$ because globally the sequence appears strongly clustered, but $L_v \sim 0$, because locally the sequence is periodic.

\subsection{The Allan Factor }

As for many temporal point processes, the Allan Factor method can be considered as one of the most appropriate measures to detect and quantify the time-clustering. Considering the timescale $\tau$ and the minimum run length $L_m$, the whole observation period can be subdivided into contiguous counting windows of duration $\tau$; thus, a discrete non-negative series of counts $\{N_k(\tau,L_m)\}$ can be obtained counting the number of runs with length $L \geq L_m $ falling in the k-th  window at that timescale $\tau$. Since the counting process representation preserves the link between the discrete time axis of $\{N_k\}$ and "real" time axis of the original point process of runs, the correlation found in $\{N_k\}$ mirrors that in the original point process \cite{Thurner1997}. The Allan Factor (AF) is a second-order measure that depends on the variability of the first difference of the counting process and its average \cite{Allan1966}:

\begin{equation}
AF_{(\tau, L_m)}=\frac{<(N_{k+1}(\tau, L_m)-N_k(\tau, L_m))^2>}{2<N_k(\tau,L_m)>}.
\end{equation}

For a homogeneous Poisson process, the AF is almost flat and fluctuates around 1 for any timescale, while any deviation from this behaviour indicates that the extremes are clustered in time. When extreme sequences are fractal, the AF behaves as a power-law with the timescale $\tau$:

\begin{equation}
AF_{(\tau, L_m)}=1+\left( \frac{\tau}{\tau_1}\right)^\alpha
\end{equation}
where $\alpha>0$ is a parameter quantifying the "strength" of the time-clusterng; $\tau_1$, known as fractal onset time, is the minimum timescale above which the fractal behavior of the point process can be significantly detected, indicating that for $\tau <<\tau_1$ the time-clustering is practically negligible \cite{Lowen1996ThePA}. The value $\alpha=0$ features Poisson point processes that are independent, uncorrelated and memoryless processes.

\section{Results}
We analyzed the point processes of extremes (or runs) of all the stations for each threshold and for $L_m$ varying from 1 to 30. Fig. 7 shows the mean interevent time versus the height for several $L_m$ and for all three thresholds.
\begin{figure}
\begin{tabular}{l}

(a)

\includegraphics[scale=0.1]{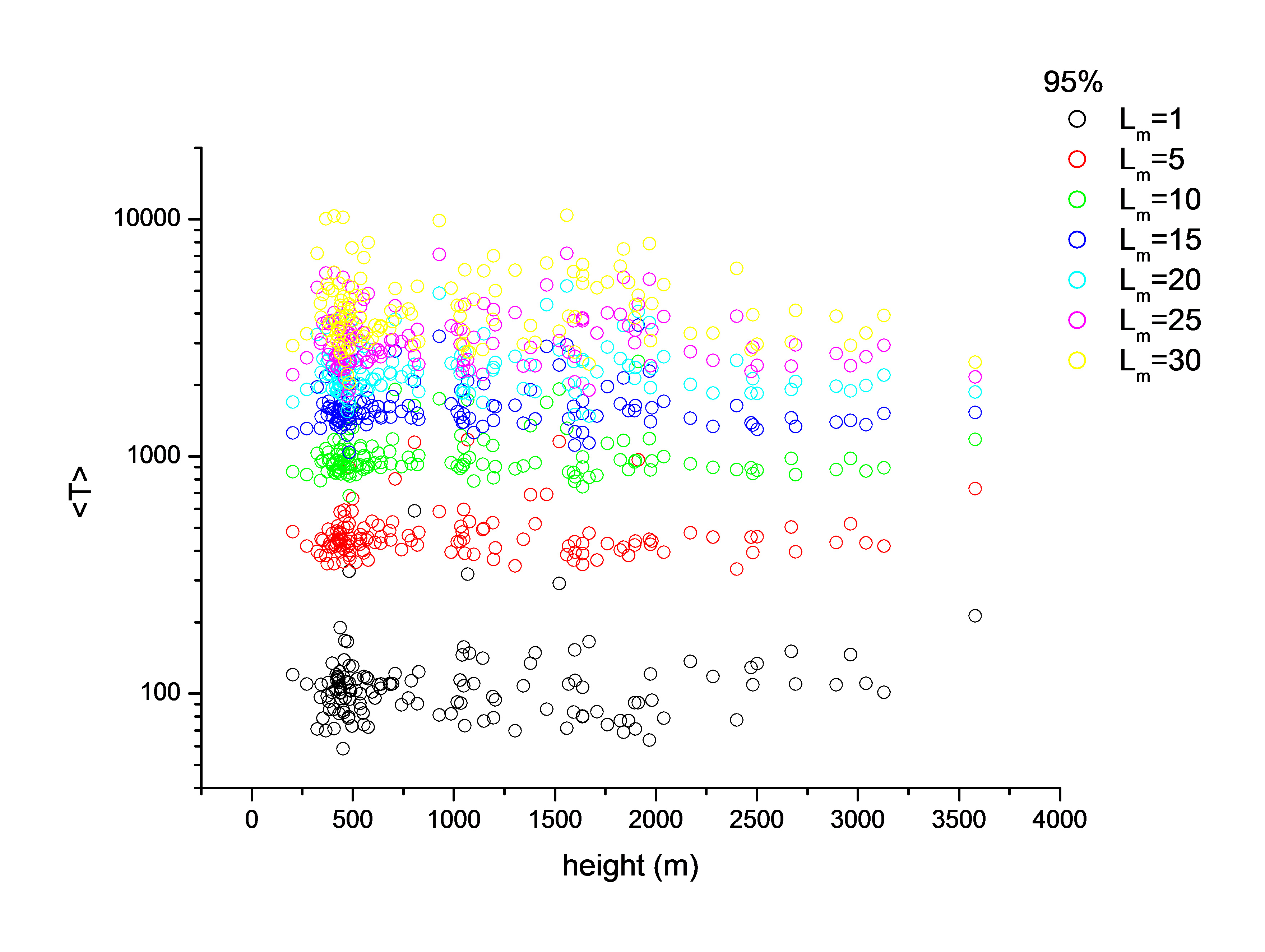} \\

(b)

\includegraphics[scale=0.1]{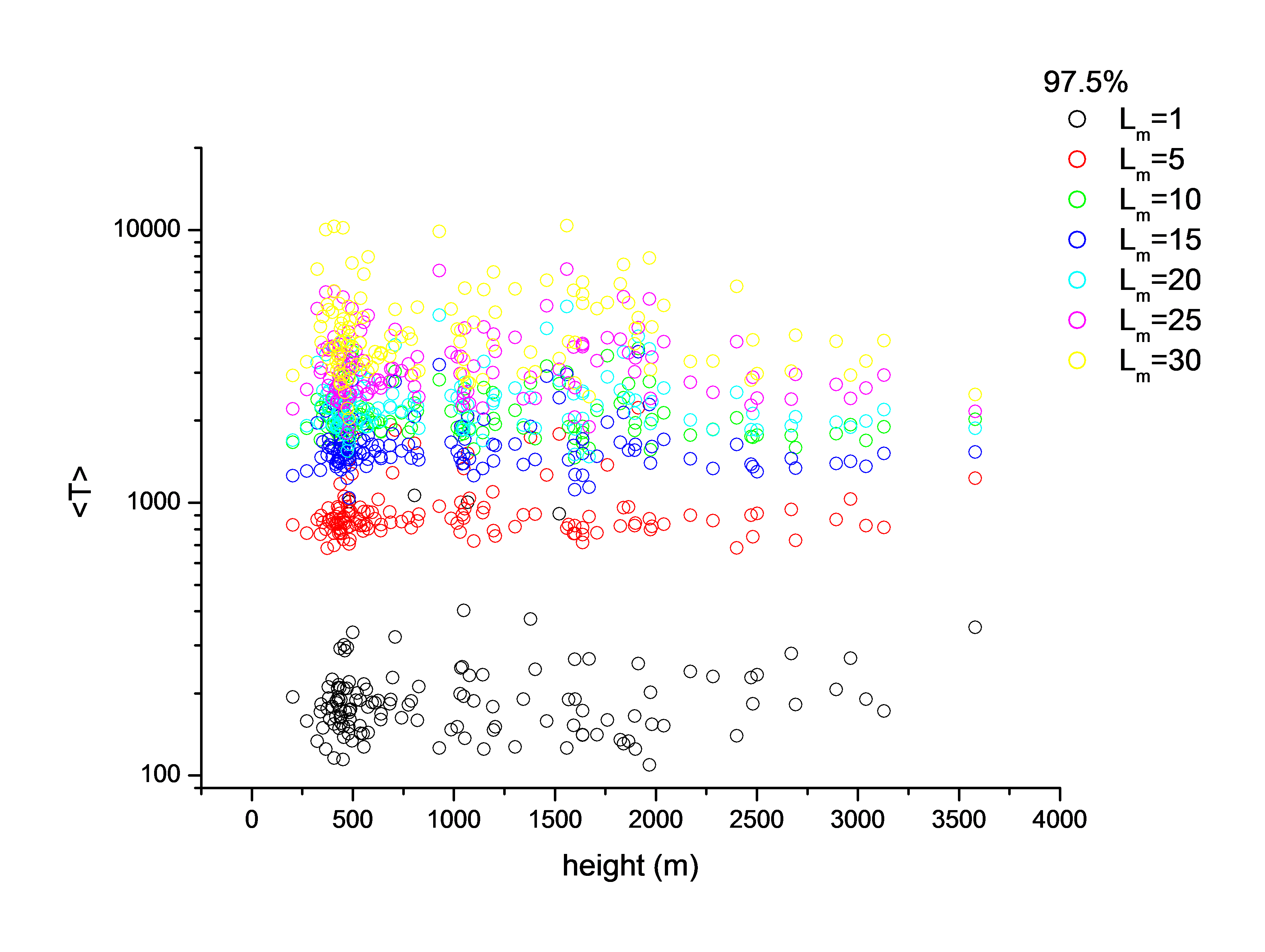}   \\

(c)

\includegraphics[scale=0.1]{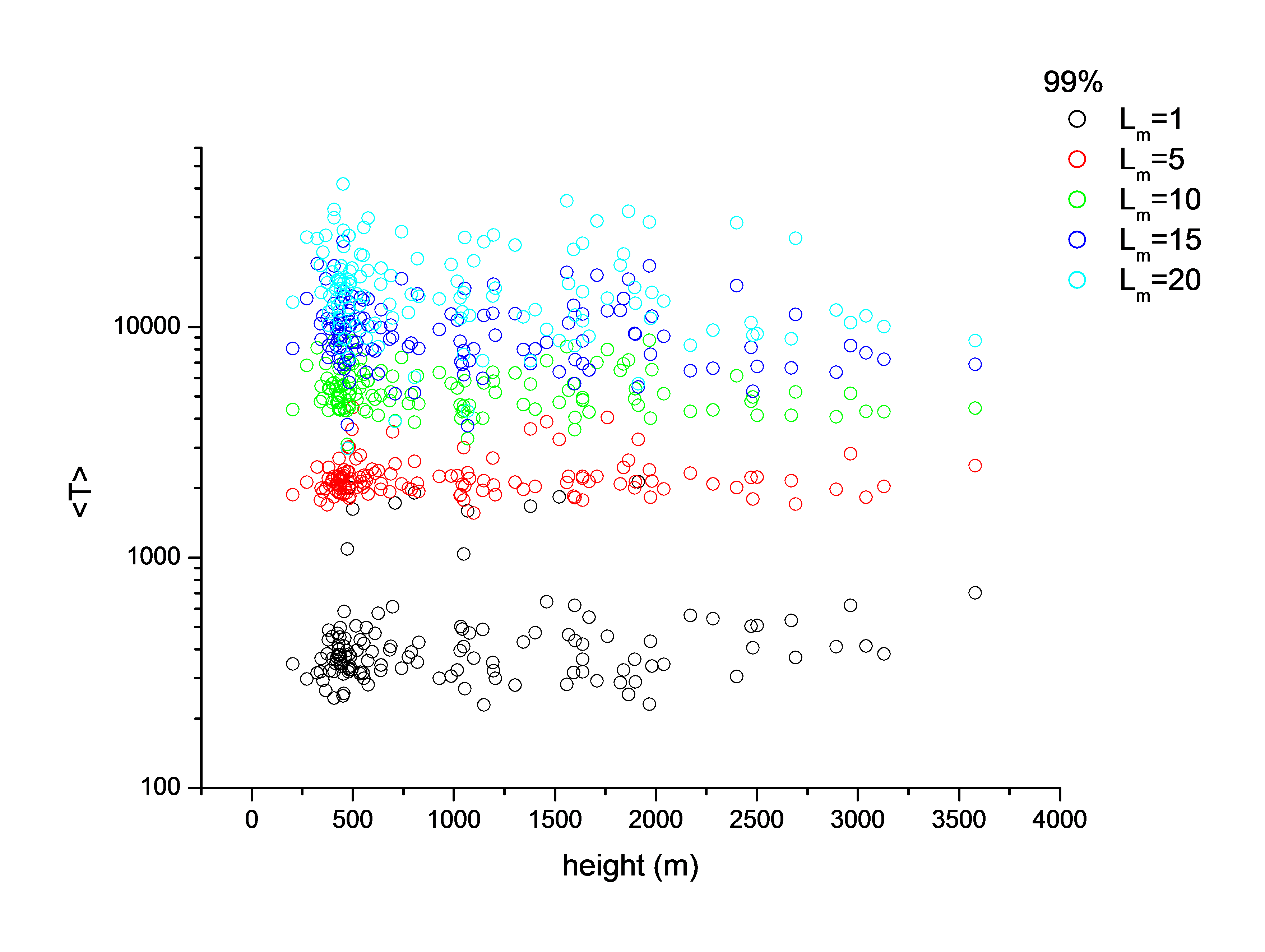} \\
\end{tabular}
\caption{Interevent time versus height for percentile-based threshold of $95\%$ (a), $97.5\%$ (b) and $99\%$ (c).}
\label{fig7}  
\end{figure}

The interevent times of wind extremes depend on $L_m$, being on average larger for large $L_m$. However, they do not show a dependence on the height at any threshold, although a slight decrease with the height can be seen for the largest $L_m$. 

Fig. 8 shows the $C_v$ and $L_v$ versus height for the three thresholds with $L_m =1$. In order to check the significance of the obtained values against random fluctuations, we calculated the $2.5th$  and $97.5th$ percentiles of the distribution of $C_v$ and $L_v$ of $1,000$ Poissonian surrogates with the same rate as the original sequence of extremes, defining as significantly time-clustered or quasi-periodic that sequences whose coefficient of variation is above the $97.5th$  or below the $2.5th$  percentile. 
\begin{figure}
\begin{tabular}{l}

(a)

\includegraphics[scale=0.1]{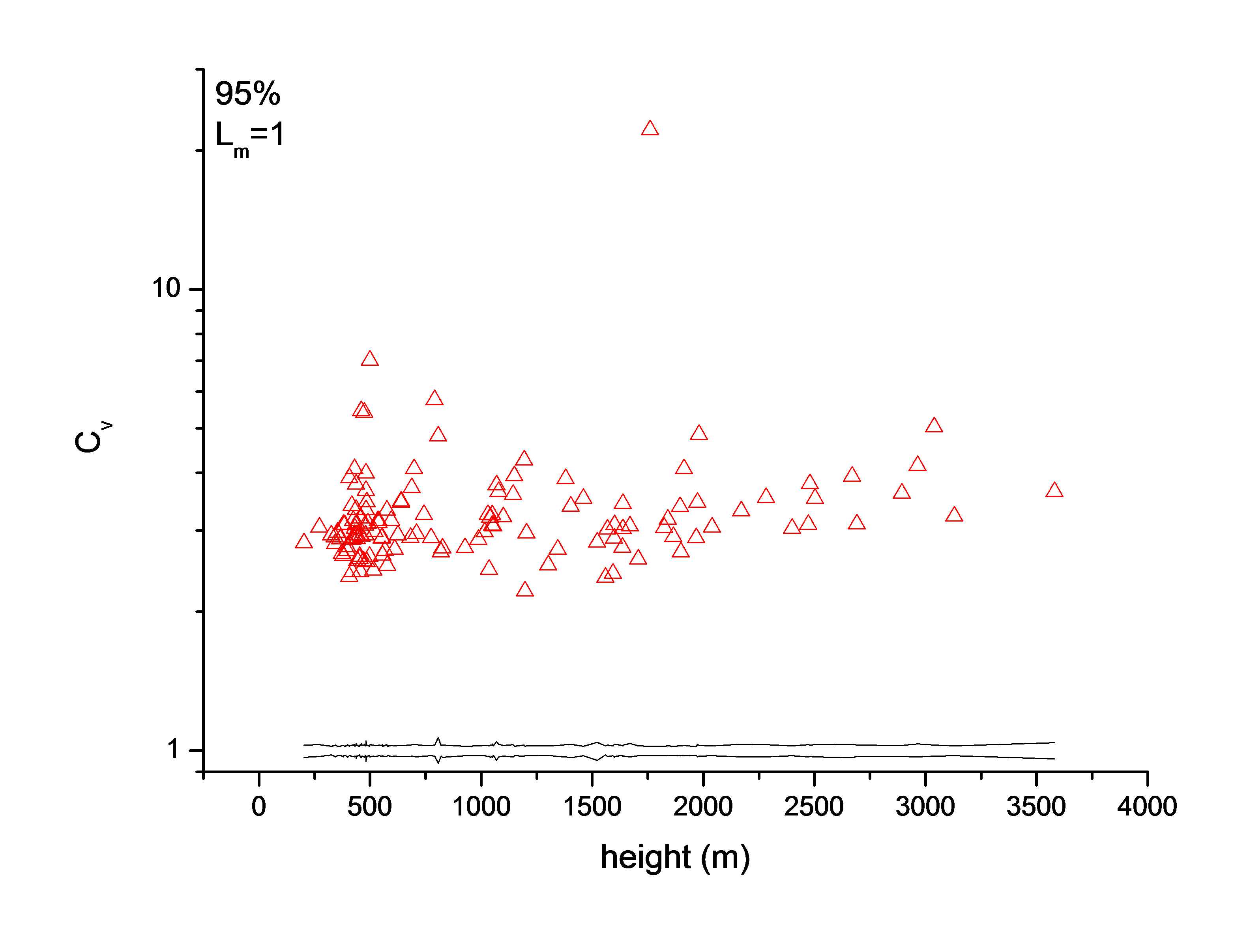} \\

(b)

\includegraphics[scale=0.1]{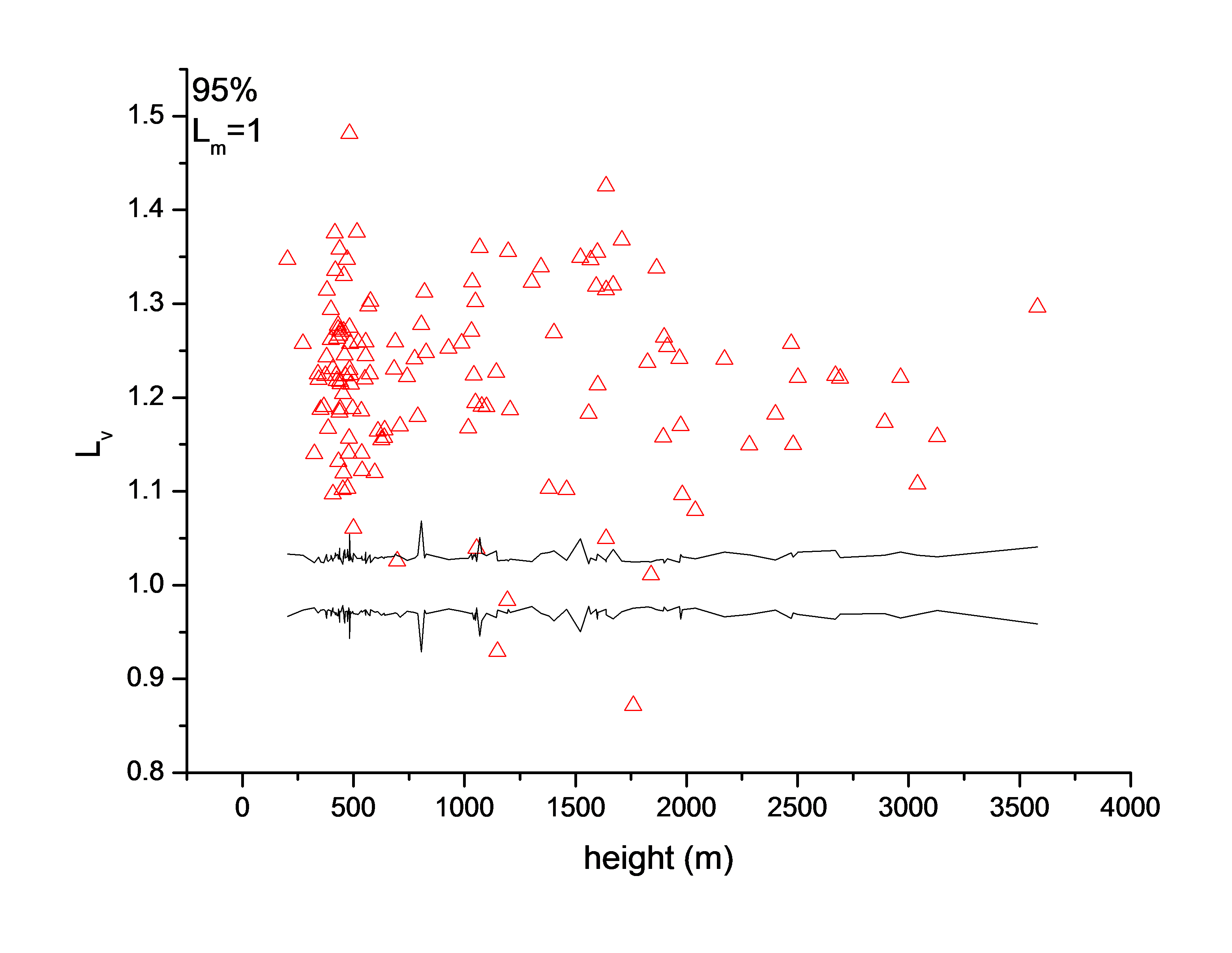}   \\

\end{tabular}
\caption{$C_v$ (a) and $L_v$ (b) for threshold of $95\%$ and $L_m =1$.}
\label{fig8}  
\end{figure}

The $C_v$ is well above the $95\%$ Poissonian confidence band, indicating that the wind extreme sequences for a threshold of $95th$ percentile of the wind speed distribution are globally clustered (Fig. 8a). At a local scale, most of the stations are clustered, although four stations show a Poissonian local behaviour and two quasi-regular (Fig. 8b). Increasing the minimum run length $L_m$, the $C_v$ still remains above the $95\%$ Poissonian confidence band, while more stations show a local Poissonian behavior of the extremes (FigSupp1). Considering the thresholds of $97.5\%$ (FigSupp2) and $99\%$ (FigSupp3), only for large $L_m$ about $20\%$ of the stations have a $C_v$ within the $95\%$ Poissonian confidence band, and about $25\%$ of the stations have $L_v$ within the $95\%$ Poissonian confidence band. Interestingly, the stations above the crossover height of about $2,000$ m a.s.l. (that we found in the relationship between the wind value and the height, Fig. 2) appear locally time-clusterized for any threshold, while many stations below $2,000$ m a.s.l.  tend to be locally Poissonian with the increase of $L_m$ at any of the three thresholds.
For each station and for each threshold we calculated the Allan Factor and its  $95\%$ confidence band obtained enveloping the $2.5t$ and $97.5th$ percentiles of $1,000$ Poissonian surrogates with the same rate as the original sequences corresponding to each timescale. 
Fig. 9 shows, as an example, the AF curve of the station AIG (381 m a.s.l.) and Fig. 10 that of GEN (1600 m a.s.l.) for the three thresholds and 
$Lm = 5$. 
Both extreme sequences are clustered and are significantly different from a Poisson process over a large range of timescales. At very large timescales, the AF of both sequences drops down, indicating a loss of clustering at those timescales. AIG seems also characterized by a fractal behaviour, since its AF is rather straight on log-log scales for a quite large range of timescales; while GEN does not show a clear evidence of fractal behaviour, because its AF curve is not straight on a quite large range of timescales. 
Increasing the minimum run length Lm, the AF tends to be closer to the $95\%$ Poisson confidence band and to loose its clustering behaviour, as it is shown, for example, by AIG in Fig. 11. At very small timescales, the AF is below the $95\%$ confidence band. This could indicate that for very small timescales up to about $6$ hours, the sequence of the extremes could be modelled by a regular or quasi-periodic process.

\begin{figure}
\begin{tabular}{l}

(a)

\includegraphics[scale=0.1]{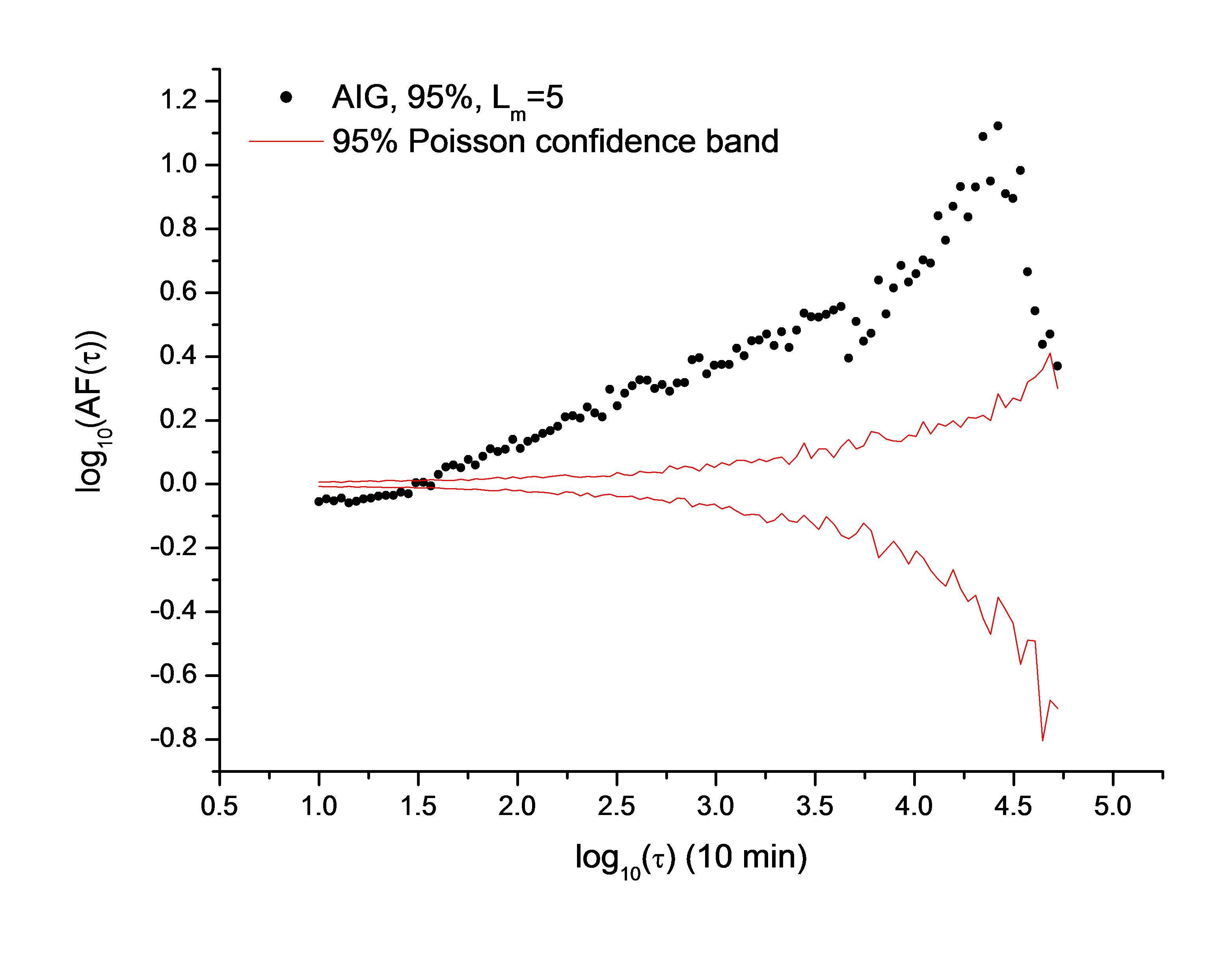} \\

(b)

\includegraphics[scale=0.1]{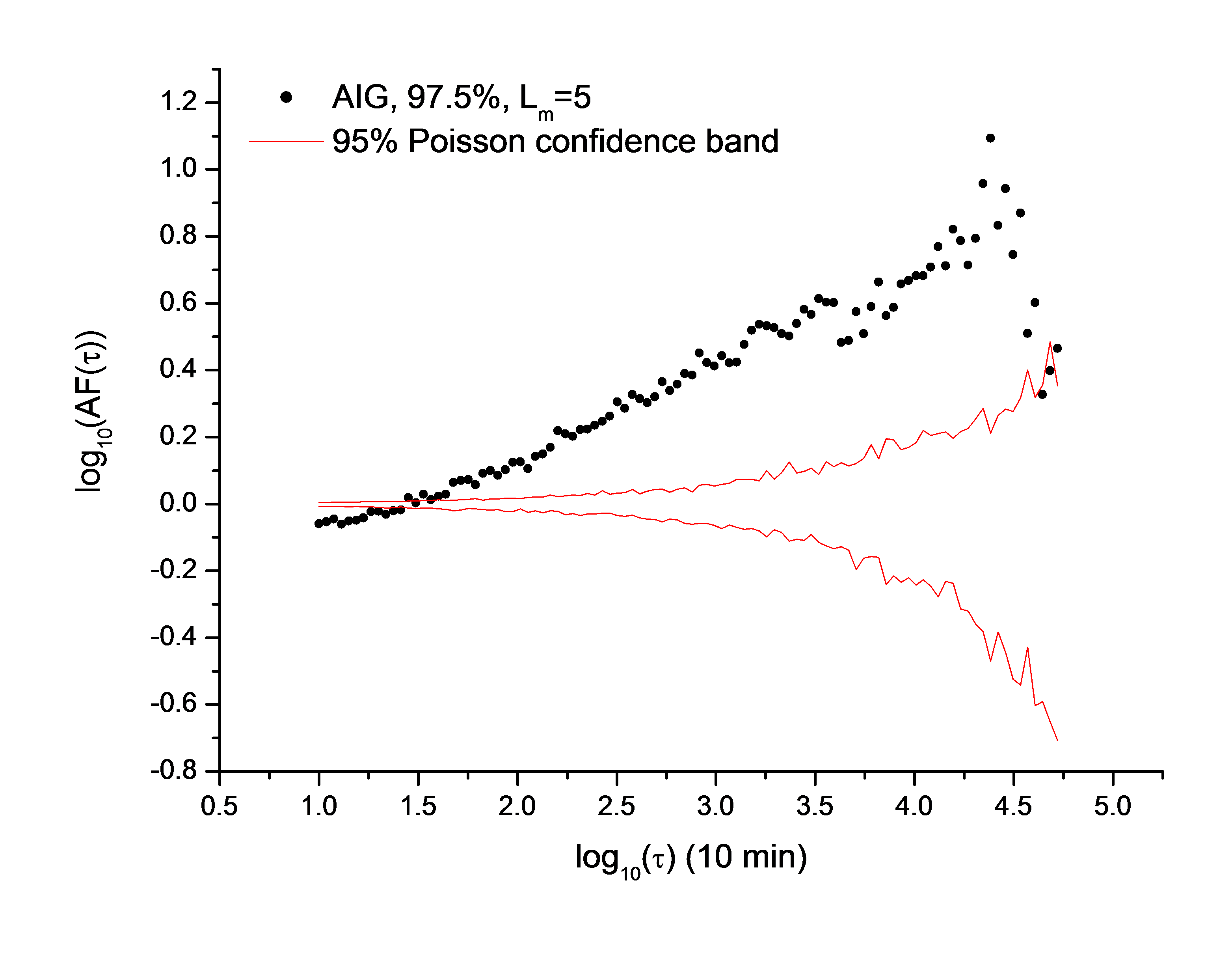}   \\

(c)
\includegraphics[scale=0.1]{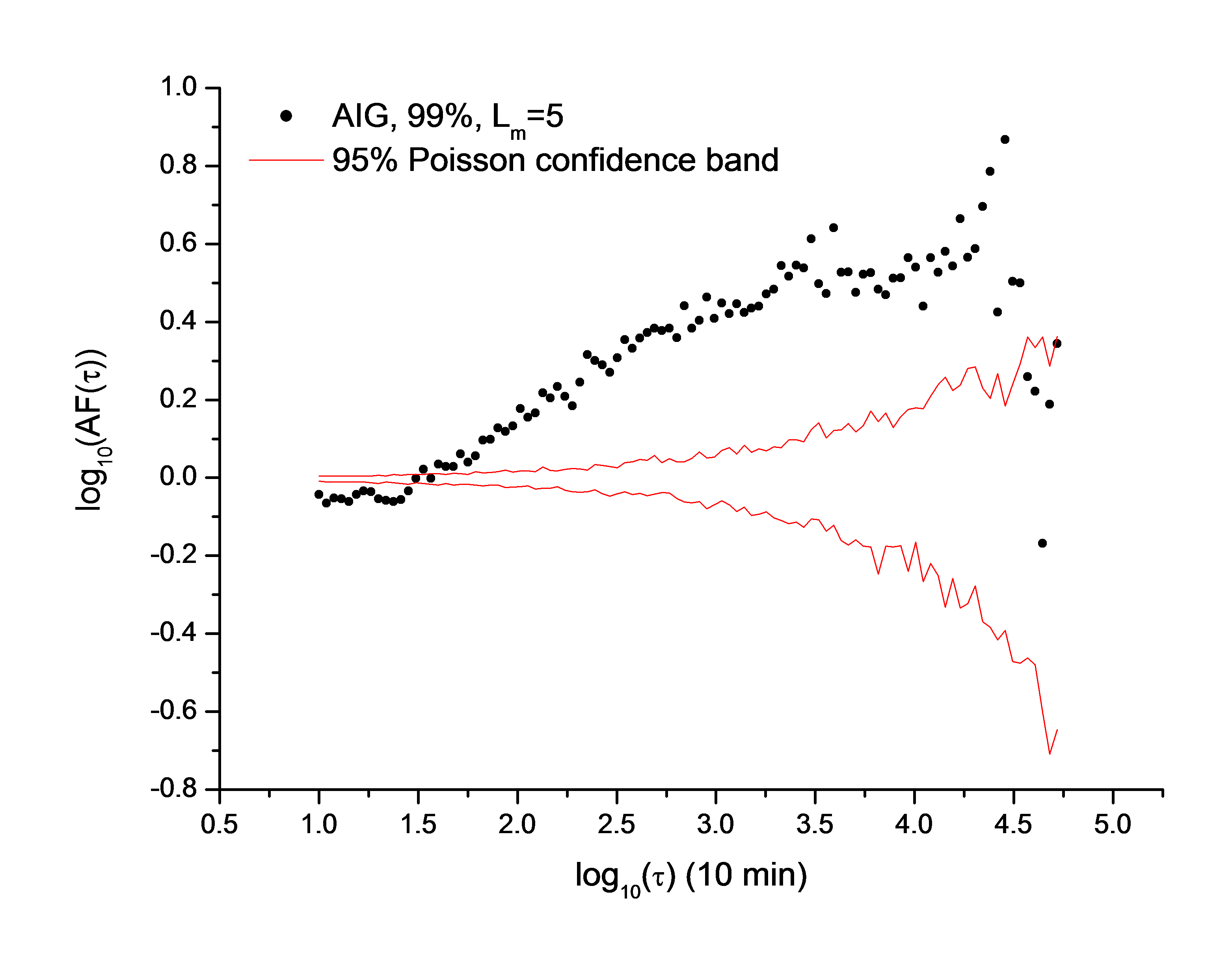}  \\
\end{tabular}
\caption{Allan Factor for station AIG for (a) $95\%$, (b) $97.5\%$, and (c) $99\%$, with $L_m=5$. Dark pink lines delimit the $95\%$ confidence band on the base of $1,000$ Poisson surrogates. }
\label{fig9}  
\end{figure}

\begin{figure}
\begin{tabular}{l}
(a)
\includegraphics[scale=0.1]{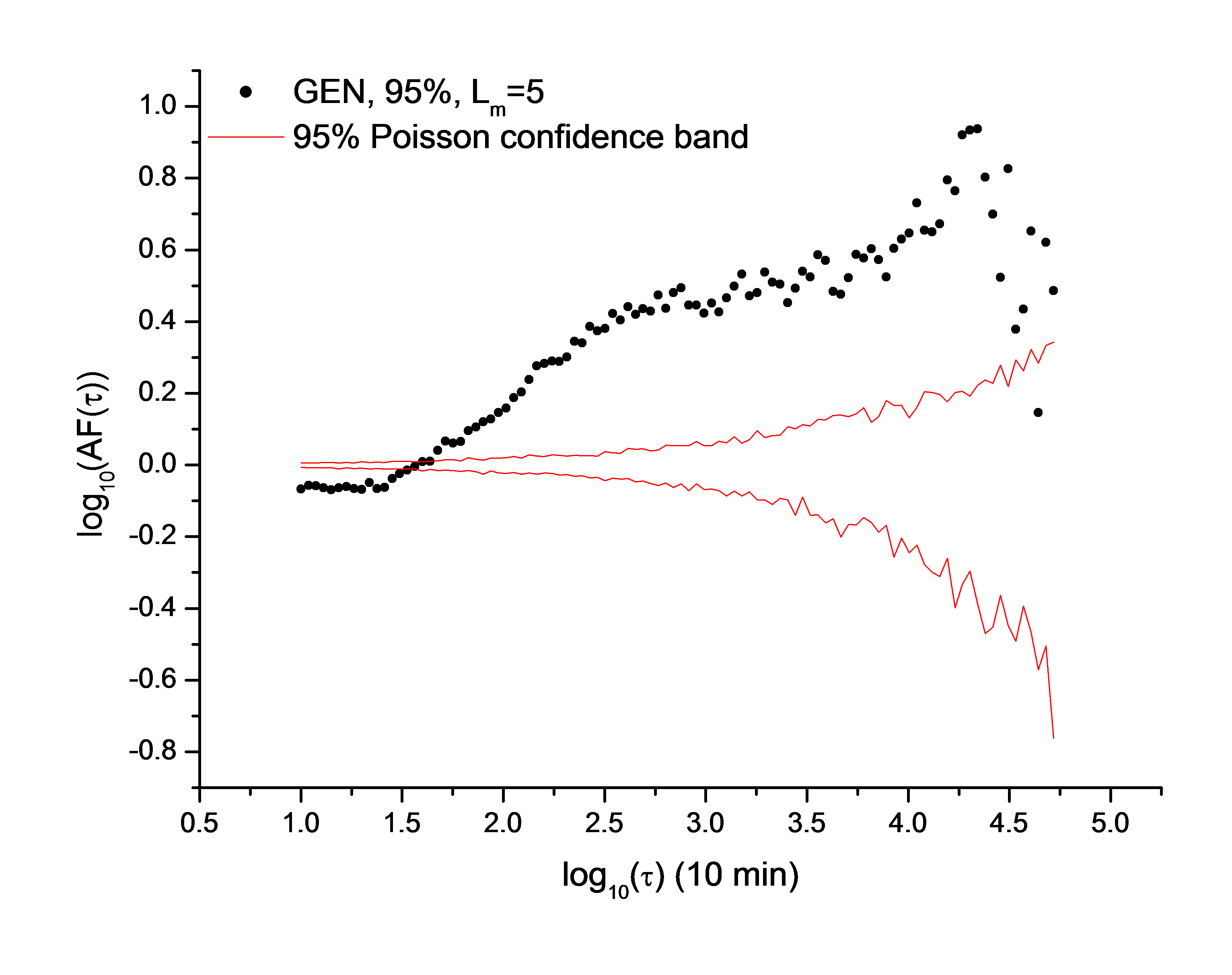} \\
(b)
\includegraphics[scale=0.1]{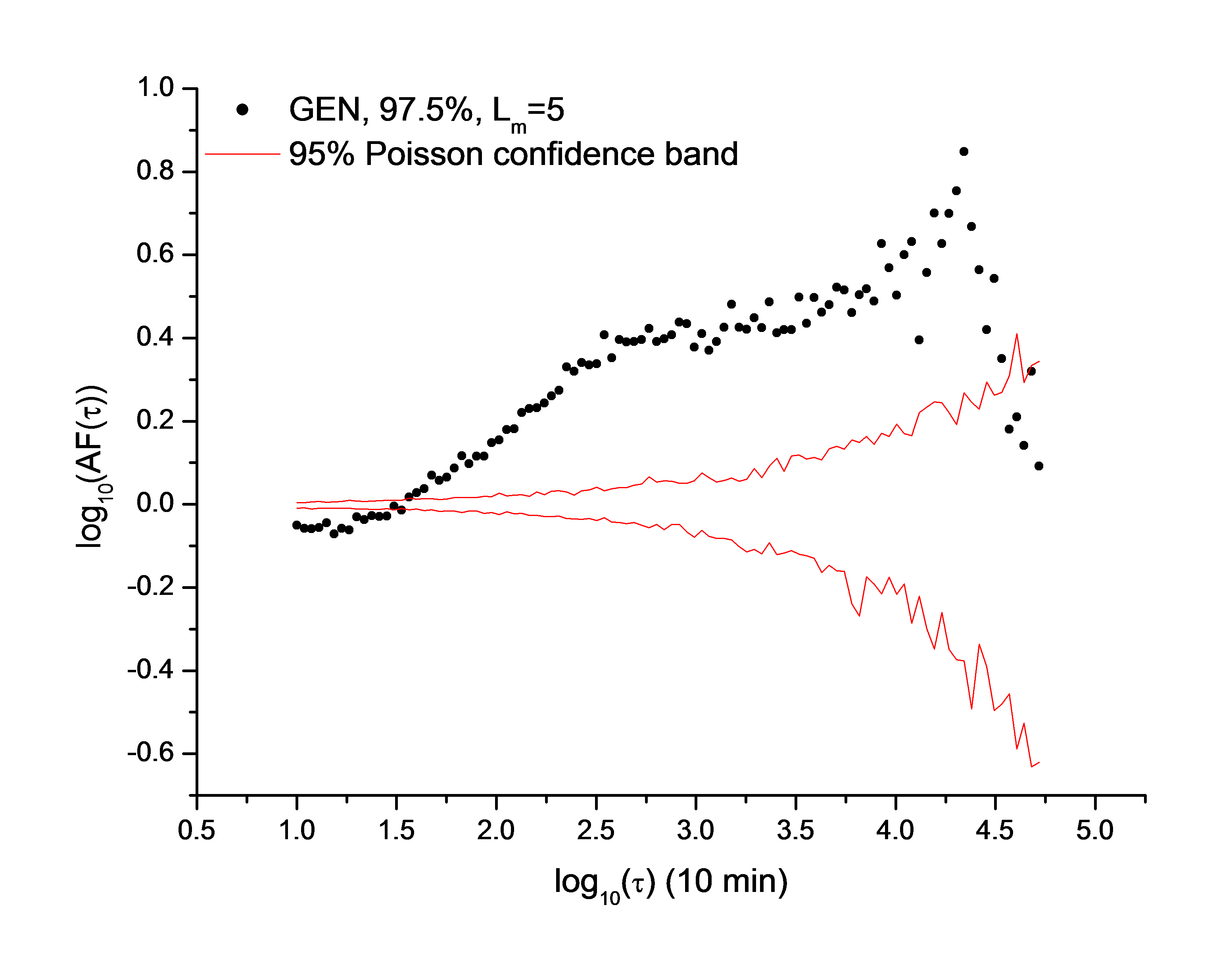}   \\
(c)
\includegraphics[scale=0.1]{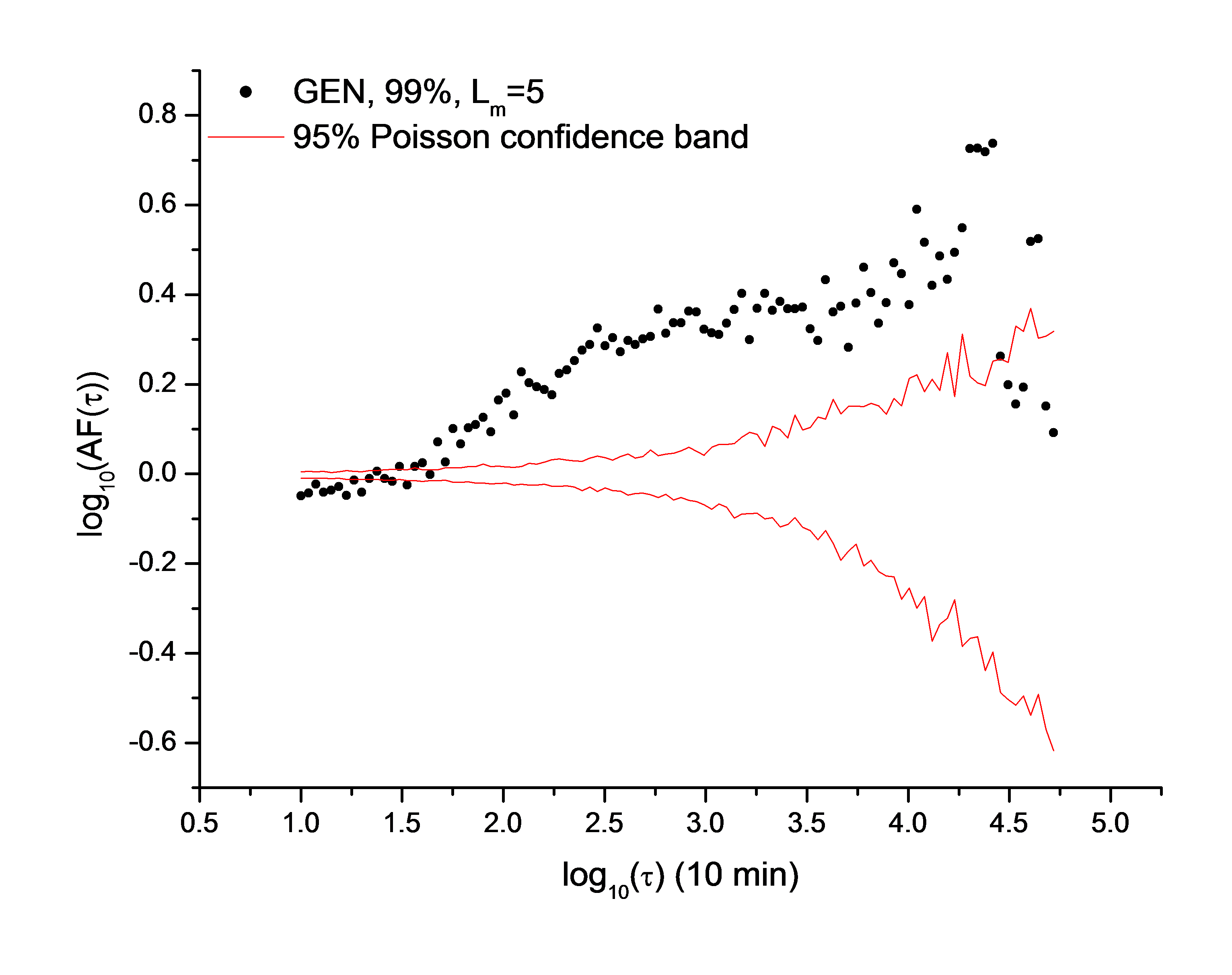} \\
\end{tabular}
\caption{Allan Factor for station GEN for (a) $95\%$, (b) $97.5\%$, and (c) $99\%$, with $L_m=5$. Dark pink lines delimit the $95\%$ confidence band on the base of $1,000$ Poisson surrogates. }
\label{fig10}  
\end{figure}

\begin{figure}
\begin{tabular}{l}
(a)
\includegraphics[scale=0.1]{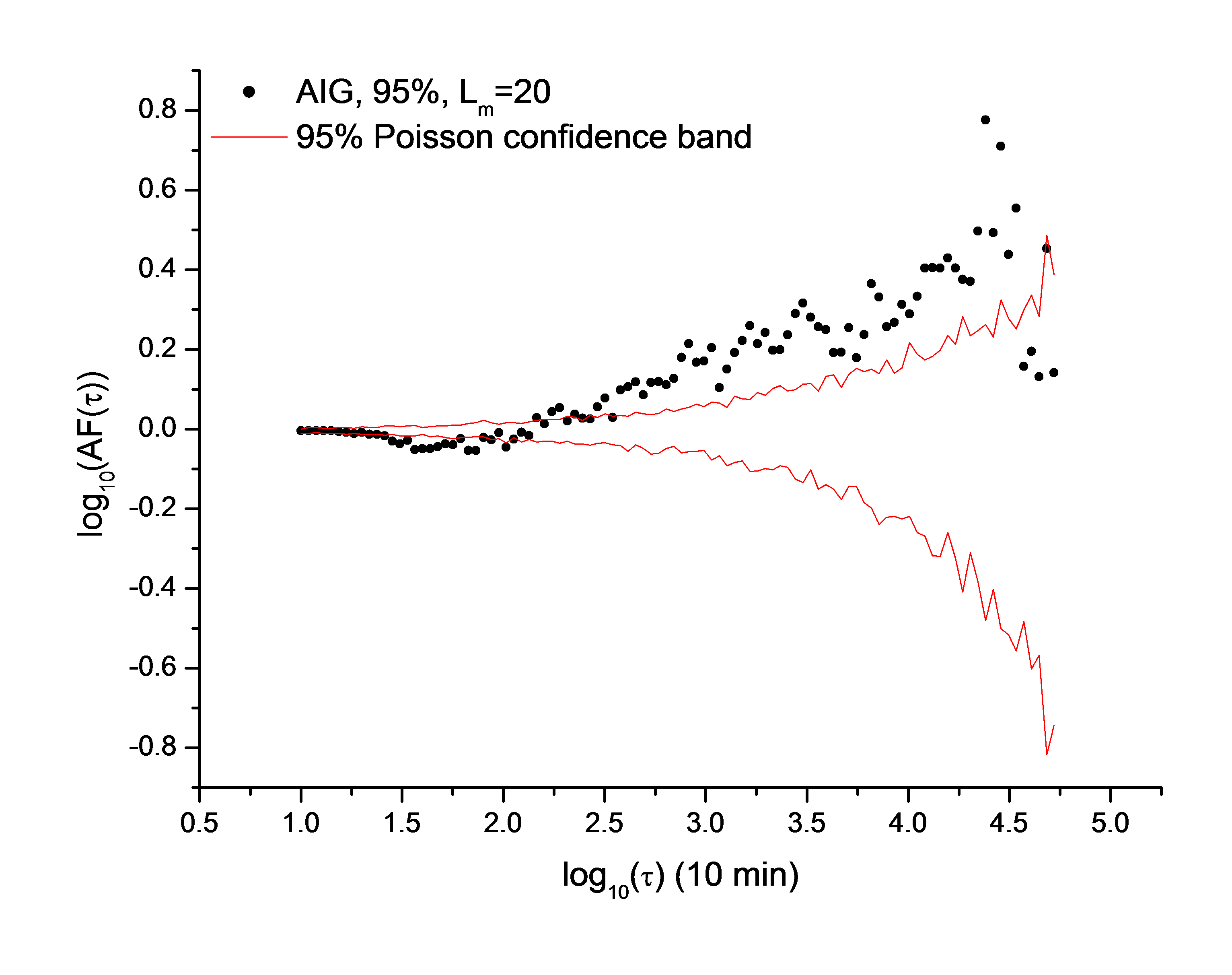} \\
(b)
\includegraphics[scale=0.1]{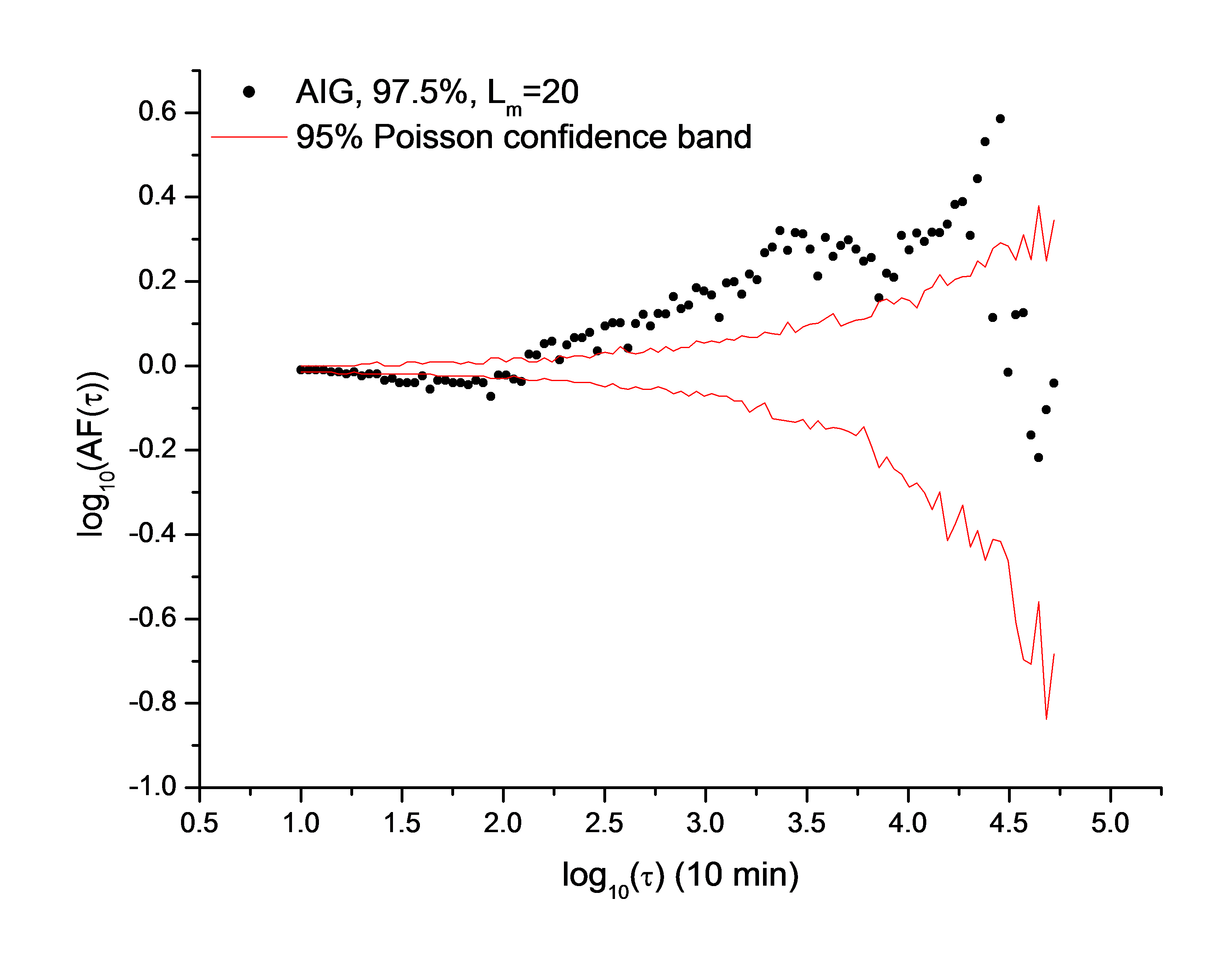}   \\
(c)
\includegraphics[scale=0.1]{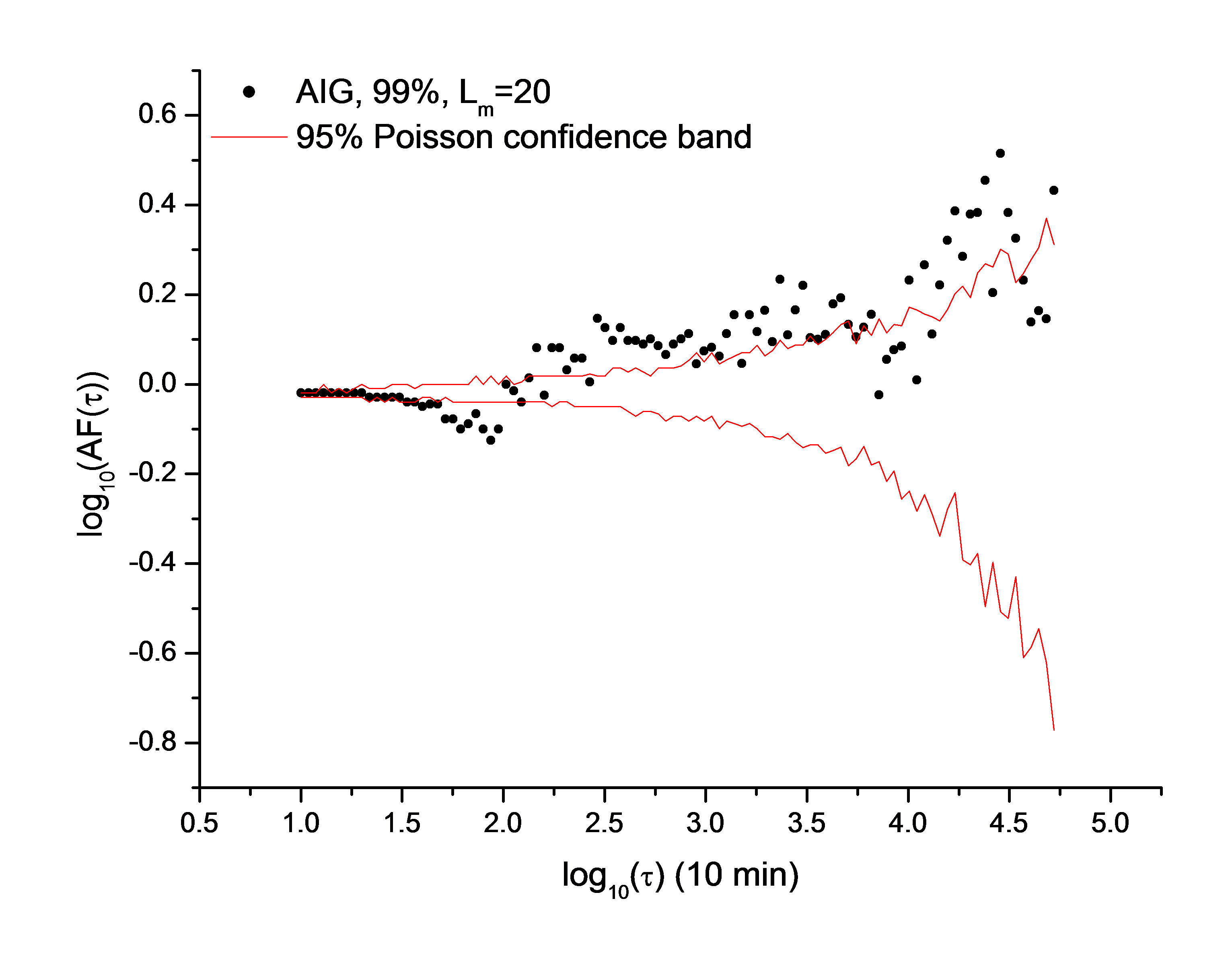} \\
\end{tabular}
\caption{Allan Factor for station AIG for (a) $95\%$, (b) $97.5\%$, and (c) $99\%$, with $L_m=20$. Dark pink lines delimit the $95\%$ confidence band on the base of $1,000$ Poisson surrogates.}
\label{fig11}  
\end{figure}

Fig. 12 shows the departure (Dp) of each station (height) from the Poissonian behaviour, calculated as the difference between the AF and the $97.5th$ curve of the Poissonian surrogates for each timescale above $200$ min (we excluded the smaller timescale range, in which the process could be considered as quasi-regular). As an example, Fig. 12 shows Dp for the threshold of $95\%$ and $L_m=1$ (Fig. 12a), threshold $97.5\%$ and $L_m =15$ (Fig. 12b) and threshold $99\%$ and $L_m =20$ (Fig. 12c).

\begin{figure}
\begin{tabular}{c}
(a)
\includegraphics[scale=0.7]{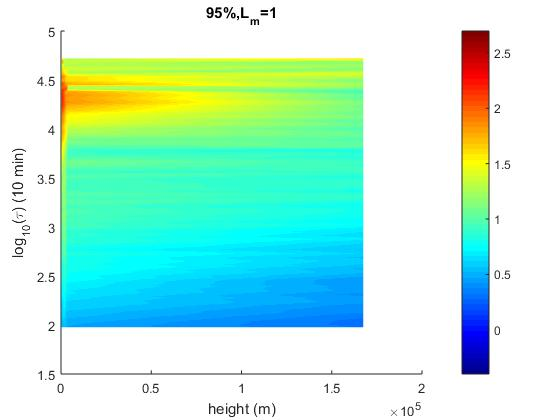} \\
(b)
\includegraphics[scale=0.7]{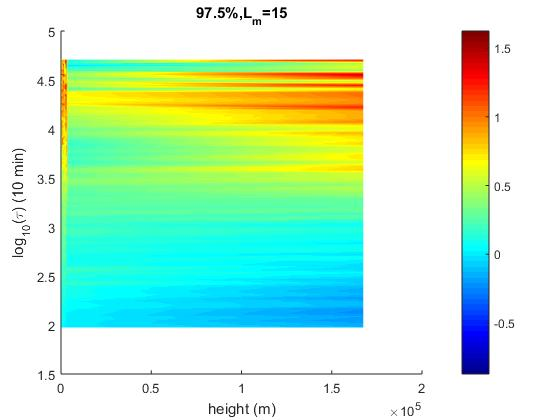}   \\
(c)
\includegraphics[scale=0.7]{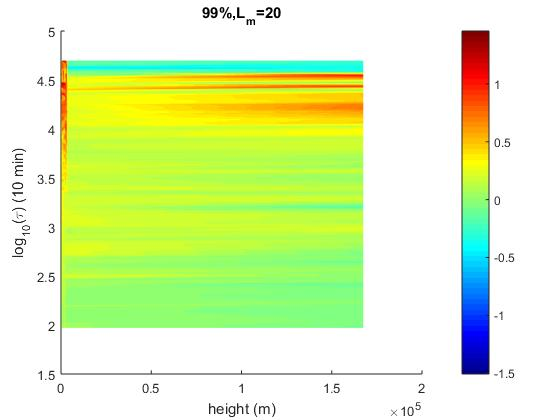} 
\end{tabular}
\caption{Departure from Poissonian behaviour of each station versus height and timescale for (a) $95\%$ and $L_m =1$, (b) $97.5\%$ and $L_m=15$, (c) $99\%$ and $L_m=20$.}
\label{fig12}  
\end{figure}

The departure $Dp$ for all the thresholds and all the minimum run length $L_m$ is shown in supplementary files FigSupp4 (for threshold of $95\%$), FigSupp5 (for threshold of $97.5\%$) and FigSupp6 (for threshold of $99\%$). Negative values of Dp at certain timescale ranges indicate that the extreme sequences are not time-clustered at those timescales, while positive values indicate the existence of time-clustering. For the threshold of $95\%$ (FigSupp4) the extremes are more time-clustered at large timescales and for any height when $L_m$ is rather small. By increasing $L_m$, the time-clustering effect involves more timescales, even lower, and seems more pronounced for large heights. For the threshold of $97.5\%$ (FigSupp5) the time-clustering phenomena is quite similar to that found for $95\%$ threshold. For the threshold of $99\%$ (FigSupp6), the time-clustering becomes less pronounced; in fact Dp generally decreases; furthermore it looks more homogeneously distributed the timescale–height space, and the relative dominance of time-clustering shown by higher stations respect to the lower ones is not discernible any more.

\section{Conclusions}

In this work the sequences of extreme events in wind speed measured by 132 stations of the MeteoSwiss weather network in Switzerland were analysed. The main findings of this studies can be summarized as follows:

\begin{enumerate}
\item for each station the wind speed extremes were defined as values above a certain percentile-based threshold (run); the duration of the extreme (or length of the run) was related to the number of consecutive values above the threshold; a crossover height of about $2,000$ m a.s.l. was found in the relationship between the wind value corresponding to the percentile of the wind speed distribution and the height;
\item the probability density function of the extreme duration shows a decreasing behaviour with the duration of the extreme;
\item although each station is characterized by its own extreme duration distribution, the average distribution of the durations of the extremes does not depend on the threshold;
\item the sequence of the wind extremes can be described by a temporal point process marked by the run length; two representations can be used: interevent times and counting process. Both representations allow to investigate the time-clustering properties of the sequences of the wind extremes;

\item the behaviour of the global coefficient of variation suggests that for $95\%$ threshold the extreme sequence of the majority of the stations are globally clustered, while increasing the threshold some stations tend to behave Poissonianly for larger $L_m$; 

\item the behaviour of the local coefficient of variation highlights the presence of the crossover height of about $2,000$ m a.s.l. found in the relationship between the percentile-based wind value and height. The stations located above such crossover height seem to be locally clustered at any threshold and for any $L_m$;

\item the Allan Factor, which is a well-known measure to detect the timescale range where time-clustering of a point process exists, shows that for  $95\%$ and $97.5\%$ thresholds the time-clustering of the extremes extends to lower timescales with the increase of $L_m$ and characterizes especially the higher stations.
\end{enumerate}

\begin{acknowledgements}

This research was partly supported by the National Research Programme 75
"Big Data" (PNR75) of the Swiss National Science Foundation (SNSF).
The authors thank MeteoSwiss for providing the data.
 L. Telesca thanks the support of the "Scientific Exchanges" project n$^\circ$ 180296 funded by the SNSF.
M. Laib thanks the support of "Soci\'et\'e Acad\'emique Vaudoise" (SAV) and the Swiss Government Excellence Scholarships.
\end{acknowledgements}

\bibliographystyle{spmpsci}      % mathematics and physical sciences

\bibliography{xampl}
%\begin{thebibliography}{}
%%
%% and use \bibitem to create references. Consult the Instructions
%% for authors for reference list style.
%%
%\bibitem{ref1}
%B. C. Ummels, M. Gibescu, E. Pelgrum, W. L. Kling and A. J. Brand, "Impacts of Wind Power on Thermal Generation Unit Commitment and Dispatch," in IEEE Transactions on Energy Conversion, vol. 22, no. 1, pp. 44-51, March 2007.
%doi: 10.1109/TEC.2006.889616
%\bibitem{ref1}
%E. Demirci, B. Cuhadaroglu,
%Statistical analysis of wind circulation and air pollution in urban Trabzon,
%Energy and Buildings,
%Volume 31, Issue 1,
%2000,
%Pages 49-53,
%ISSN 0378-7788,
%doi: 10.1016/S0378-7788(99)00002-X.
%\bibitem{refA}
%\end{thebibliography}

\end{document}